\documentclass[aps,prd,10pt,superscriptaddress,twocolumn,nofootinbib]{revtex4-1}
\usepackage{graphicx, epsfig}
\usepackage{amsmath,amssymb,amsfonts,dsfont,mathrsfs,amsthm,mathtools}
\usepackage{bm}
\usepackage{color}
\usepackage{physics}
\usepackage[usenames]{xcolor}
\usepackage{hyperref}
\usepackage{siunitx}
\hypersetup{colorlinks=true,urlcolor=blue,linkcolor=blue,citecolor=blue,filecolor=blue}
\usepackage[normalem]{ulem}
\usepackage{array}
\usepackage{hyperref}
\usepackage{booktabs}
\usepackage{cancel}
\usepackage{blindtext}
\usepackage{tikzsymbols}
\usepackage[paperwidth=625pt,top=80pt,right=60pt,bottom=80pt,left=60pt,headheight=50pt]{geometry}
\usepackage{lipsum}
\usepackage{orcidlink}

\newcommand*{\diag}{\operatorname{diag}}

\newcommand{\A}{\mathcal{A}}

\newcommand{\F}{\mathcal{F}}

\begin{document}

\title{(Super-)renormalizable hairy meronic black holes}
\begin{abstract}
We construct an analytical black hole solution in the Einstein-Maxwell-Yang-Mills theory with a  conformally coupled scalar field in four dimensions, which generalizes the charged Mart\'inez-Troncoso-Zanelli (MTZ) black hole in the presence of self-gravitating non-Abelian gauge fields. The internal gauge group is determined by the horizon curvature, becoming $SU(N)$ in the case of positive curvature and $SU(N-1,1)$ when the curvature is negative. Moreover, this solution is employed as a conformal seed to obtain new meronic spacetimes dressed with all (super-)renormalizable contributions of the scalar field, which provides the generalization of the Anabal\'on-Cisterna (AC) solution when self-gravitating non-Abelian gauge fields are included. Finally, we consider the non-Noetherian extension of the conformal scalar fields, which still yields a second-order conformally invariant scalar equation, even though the action is not. In that case, we show that static black hole solutions can also be charged with Yang-Mills fields. 
\end{abstract}

\author{Luis \surname{Avil\'es}\,\orcidlink{0000-0002-2356-5747}
}
\email{luaviles@unap.cl}
	\affiliation{Instituto de Ciencias Exactas y Naturales, Universidad Arturo Prat, Playa Brava 3256, 1111346, Iquique, Chile}
	\affiliation{Facultad de Ciencias, Universidad Arturo Prat, Avenida Arturo Prat Chac\'on 2120, 1110939, Iquique, Chile}

\author{Borja \surname{Diez}\,\orcidlink{0009-0004-3805-4036}
}
\email{borja.diez@cinvestav.mx}
\affiliation{Departamento de F\'{\i}sica, Cinvestav, Av.~IPN 2508, 07360, CDMX, M\'exico}

\maketitle
\section{Introduction}

In modern physics, Yang-Mills theory stands as one of the most successful theoretical frameworks, providing the gauge structure underlying quantum chromodynamics and the electroweak sector of the Standard Model. Beyond the perturbative regime, non-Abelian gauge fields exhibit nontrivial configurations that establish profound links between topology and quantum field theory. When coupled to gravity, Yang-Mills fields acquire additional significance, as they are known to evade the no-hair conjecture~\cite{Bizon:1990sr}. Within $SU(2)$ Yang-Mills theory, several hairy black hole solutions have been constructed (see for instance~\cite{Smoller:1997qr,Winstanley:2008ac,Volkov:1998cc,Bartnik:1988am}). However, for some of these configurations, the gauge potential effectively reduces to an Abelian sector of the theory~\cite{Bizon:1990sr,Smoller:1997qr}, since it involves only a single nontrivial component along the generators of $SU(2)$. Consequently, such solutions do not correspond to genuinely non-Abelian field configurations. Black hole solutions have also been obtained for higher-rank gauge groups, such as $SU(N)$, using numerical methods~\cite{Shepherd:2015dse,Shepherd:2016ily}.

Among all the solitonic solutions admitted by Yang-Mills theory, merons stand out as particularly important configurations. First discovered in Ref.~\cite{deAlfaro:1976qet}, they are described by gauge fields that are proportional to a pure-gauge one, that is, $A=c \,U^{-1}\dd U$, where $U(x)$ is an element of a non-Abelian gauge group, and $c$ is a constant different from zero or one. While in Abelian theories (such as Maxwell electrodynamics) this proportionality would imply that the field is itself pure gauge, in Yang-Mills theories this is no longer the case. In this sense, merons constitute genuinely non-Abelian field configurations\footnote{Recently it was shown that these configurations arise in a broad class of non-Abelian gauge theories beyond Yang-Mills, thus exhibiting a universal character~\cite{Diez:2026bmp}.}. A distinctive property of merons is that they carry half a unit of topological charge, which is localized and therefore unobservable in isolation on flat spacetime, where merons are singular. Despite this, they are believed to play a relevant role in the mechanism of confinement (see Refs.~\cite{Callan:1977qs,Callan:1977gz,Callan:1978bm,Negele:1998ev,Steele:2000xk,Polyakov:1976fu,Actor:1979in} for further details). However, when the theory is coupled to general relativity, the gravitational backreaction can cloak the singularity behind a black hole horizon, potentially making such configurations physically realizable~\cite{Canfora:2012ap,Canfora:2017yio,Canfora:2018ppu,Canfora:2022nso,Flores-Alfonso:2020ayc}. The simplicity of these intrinsically non-Abelian configurations allowed the construction, in Ref.~\cite{Canfora:2012ap}, of an intrinsically non-Abelian black hole solution within the $SU(2)$--Einstein--Yang-Mills system. The strategy relied on employing the generalized hedgehog Ansatz~\cite{Canfora:2013hedgehog,Canfora:2013maeda,Canfora:2013crystals,Chen:2014kink,Canfora:2014multisoliton,Canfora:2014cosmo,Canfora:2014monopoles,Canfora:2015sun,Chen:2016domain,AyonBeato:2016skyrmions,Canfora:2015bps,Canfora:2016orientational,Tallarita:2017popcorn,Canfora:2017einsteinskyrme,Canfora:2017sigma,Canfora:2017merons,Canfora:2025roy,Canfora:2018ppu,Canfora:2017yio,Ipinza:2020xgc,Flores-Alfonso:2020ayc,Bahamonde:2026bvh}, which provides a consistent reduction of the field equations while preserving the genuinely non-Abelian character of the gauge configuration.

On the other hand, the inclusion of scalar fields in non-Abelian gauge theories is of fundamental importance, as exemplified by the Higgs mechanism, which implements spontaneous symmetry breaking and determines the mass spectrum of the gauge sector. Since classical Yang-Mills theories are conformally invariant, the introduction of conformally coupled scalar fields arises as a natural extension that preserves the symmetry properties of the matter sector, particularly when gravity is taken into account. In gravitational physics, scalar fields also play a prominent role, and scalar-tensor theories have proven to provide a fertile framework for evading standard no-hair theorems (see Ref.~\cite{Herdeiro:2015waa} for a review). In particular, when gravitational theories are supplemented with conformally coupled scalar fields, a broad class of solutions has been constructed~\cite{Bekenstein:1974sf,Martinez:2002ru,Martinez:2005di,Charmousis:2009cm,Bardoux:2013swa,Astorino:2014mda,Anabalon:2009qt,Cisterna:2021xxq,Caceres:2020myr,Barrientos:2022avi,Aviles:2018vnf,Cisterna:2023uqf,Bravo-Gaete:2025vyd,Barrientos:2023tqb}, including gravitational instantons~\cite{Arratia:2020hoy,Corral:2025npd,deHaro:2006ymc}. In asymptotically flat spacetime and in the absence of a self--interacting scalar potential, there exists an analytic black hole solution originally found by Bekenstein~\cite{Bekenstein:1974sf}, and independently by Bocharova, Bronnikov, and Melnikov~\cite{Bocharova:1970}, commonly known as the BBMB black hole, which has been somewhat controversial~\cite{Sudarsky:1997te}, since the scalar field diverges at the event horizon. However, when a cosmological constant is included together with a conformal scalar potential, another analytic black hole solution does exist. Such a solution was found by Mart\'{i}nez, Troncoso, and Zanelli in Ref.~\cite{Martinez:2002ru}, and will be referred to hereafter as the MTZ solution. This geometry can be regarded as the (anti)de Sitter analog of the BBMB solution, in which the singularity of the scalar field is now hidden behind the event horizon. The electrically charged extension of this solution, sourced by Maxwell fields, as well as its extension to other horizon topologies, was reported in Ref.~\cite{Martinez:2005di}. Moreover, the integrability properties of this theory have been extensively investigated, leading to a wide variety of exact solutions. These include configurations endowed with Newman-Tamburino-Unti (NUT) charge~\cite{Bardoux:2013swa}, accelerating solutions~\cite{Charmousis:2009cm,Astorino:2013xxa}, and rotating geometries~\cite{Astorino:2014mda}. In addition, it has been shown that the full Pleba\'{n}ski-Demia\'{n}ski family provides an exact class of solutions to this system~\cite{Anabalon:2009qt,Cisterna:2021xxq}. The renormalization of this theory was studied in Ref.~\cite{Anastasiou:2022wjq} by restoring the on-shell conformal symmetry in the bulk. In dimensions higher than four, this theory does not admit black hole solutions~\cite{Xanthopoulos:1992fm,Klimcik:1993cia}, and the inclusion of conformal higher-curvature corrections becomes necessary in order to construct them~\cite{Oliva:2011np,Giribet:2014bva,Chernicoff:2016jsu,Babichev:2023rhn}. In this context, motivated by the interplay between scalar fields and non-Abelian gauge dynamics, we show that the MTZ black hole can be extended by incorporating non-Abelian gauge fields and allowing for horizons of nonvanishing constant curvature. 

Furthermore, a large family of analytic solutions has been studied when conformal symmetry at the level of the action is relaxed, either by allowing for a generic nonminimal coupling parameter or by introducing scalar self-interaction potentials~\cite{Ayon-Beato:2004nzi,Ayon-Beato:2005yoq,Anabalon:2012tu,Zhao:2013isa,Xu:2014uha,Ayon-Beato:2015ada,Fan:2015tua,Erices:2017izj,Ayon-Beato:2005pnc,Ayon-Beato:2015mxf,Erices:2024iah}. A particularly interesting case arises when all power-counting super-renormalizable contributions to the scalar potential are taken into account, as this leads to a wide range of spacetime configurations~\cite{Anabalon:2012tu}. Moreover, in Ref.~\cite{Ayon-Beato:2015ada} was shown that these solutions can be generated through a conformal transformation of the fields starting from the Einstein equations with a cosmological constant, coupled to a conformally coupled scalar field with a conformal potential. This mapping can be straightforwardly extended to include Maxwell fields, since the theory is conformally invariant in four dimensions. In this work, we extend these transformations by incorporating non-Abelian gauge fields in four dimensions, where Yang-Mills theory also exhibits conformal invariance.

The manuscript is organized as follows. In Sec.~\ref{sec:the-theory}, we present the Einstein-Maxwell-Yang-Mills theory with a cosmological constant and a conformally coupled scalar field, and we fix our notation. In Sec.~\ref{sec:meronic-black-hole-solution}, we introduce a new dyonic black hole solution dressed with scalar fields and self-gravitating gauge fields, which generalizes the MTZ black hole. Some of its properties and interesting limits are discussed. Then, in Sec.~\ref{sec:super-renormalizable}, we extend the solution-generating technique introduced in Ref.~\cite{Ayon-Beato:2015ada} to include non-Abelian gauge fields. Using the meronic MTZ black hole as a conformal seed, we construct a new solution supported by Yang-Mills fields and a scalar field endowed with a (super-)renormalizable self-interaction potential. Finally, in Sec.~\ref{sec:nN-meronic-BH}, we explore the non-Noetherian extension of conformally coupled scalar fields~\cite{Fernandes:2021dsb,Ayon-Beato:2023bzp}, generalizing the solutions presented in Refs.~\cite{Fernandes:2021dsb,Ayon-Beato:2024vph} to the case of non-Abelian gauge fields. We conclude in Sec.~\ref{sec:discussion} with a discussion of our main results and possible directions for future work.

\section{The theory}\label{sec:the-theory}
In this work, we focus on the Yang-Mills theory coupled to Einstein-Maxwell gravity with a cosmological constant $\Lambda$ and to a conformally coupled scalar field. The dynamics of the system is described by the following action principle
\begin{equation}\label{action}
	I[g_{\mu \nu},\A_{\mu} ,A_\mu,\phi]=I_{\rm G} + I_{\rm Max}+I_{\rm YM}+I_{\rm SF}\,,
\end{equation}
where
\begin{subequations}\label{I's}
	\begin{align}
  I_{\rm G}&=\frac{1}{16\pi G}\int_{\mathcal{M}}\dd^4x\sqrt{|g|}\left(R-2\Lambda\right)\,,\\
  I_{\rm Max}&=-\frac{1}{4}\int_{\mathcal{M}}\dd^4x\sqrt{|g|}\F_{\mu\nu}\F^{\mu\nu}\,,\\
  I_{\rm YM}&=\frac{1}{2}\int_\mathcal{M}\dd^4x\sqrt{|g|}\Tr(F_{\mu\nu}F^{\mu\nu})\label{I-YM}\,,\\
  I_{\rm SF}&=-\int_\mathcal{M}\dd^4x\sqrt{|g|}\left(\frac{1}{2}(\nabla\phi)^2+\frac{1}{12}R\phi^2+\lambda\phi^4\right)\label{I-f}\,.
\end{align}
\end{subequations}
Here, $G$ and $\Lambda$ are the gravitational and cosmological constants, respectively, $\lambda$ is the coupling of the conformal potential, and $(\nabla\phi)^2:=\nabla_\mu\phi\nabla^\mu\phi$ corresponds to the kinetic term for the scalar field. From now on, Latin indices will denote $SU(2)$ indices, while Greek indices will refer to spacetime indices. The non-Abelian field strength is given in terms of the non-Abelian gauge connection in the adjoint representation of the group, i.e.,  $A_\mu=A_\mu^{a}t_a$, according to
\begin{equation}\label{F-YM}
	F_{\mu\nu}=F_{\mu\nu}^{a}t_a=\partial_\mu A_\nu-\partial_\nu A_\mu +e[A_\mu,A_\nu]\,,
\end{equation}
where $e$ corresponds to the Yang-Mills coupling, while the Abelian field strength is given by
\begin{equation}
    \F_{\mu\nu}=\partial_\mu \A_\nu- \partial_\nu \A_\mu\,.
\end{equation}

The field equations of the theory are obtained by performing arbitrary variations of the action with respect to the metric, the Maxwell and Yang-Mills gauge potentials, and the scalar field, yielding
\begin{subequations}\label{EOM}
	\begin{align}
  G_{\mu\nu}+\Lambda g_{\mu\nu}&=8\pi G\left(T^{(\rm Max)}_{\mu\nu}+T^{(\rm YM)}_{\mu\nu}+T^{(\rm SF)}_{\mu\nu}\right)\,,\label{EOM-g}\\
 \nabla_\mu \F^{\mu\nu}&=0\label{eom-Maxwelll}\,,\\
 \nabla_\mu F^{\mu\nu}+e[A_\mu,F^{\mu\nu}]&=0\label{EOM-A}\,,\\
  \Box\phi&=\frac{1}{6}R\phi+4\lambda\phi^3\label{EOM-f}\,,
\end{align}
\end{subequations}
respectively, where $G_{\mu\nu}:=R_{\mu\nu}-\frac{1}{2}g_{\mu\nu}R$ denotes the Einstein tensor, and we have defined the stress--energy momentum tensor for the Maxwell, Yang-Mills and scalar fields, as
\begin{subequations}\label{Tmunu}
	\begin{align}
    T^{(\rm Max)}_{\mu\nu}&=\F_{\mu\lambda}\F_{\nu}^{~\lambda}-\frac{1}{4}g_{\mu\nu}\F_{\alpha\beta}\F^{\alpha\beta}\,,\\
  T^{(\rm YM)}_{\mu\nu}&=-2\Tr\left(F_{\mu\lambda}F_{\nu}^{~\lambda}-\frac{1}{4}g_{\mu\nu}F_{\alpha\beta}F^{\alpha\beta}\right)\,,\label{Tmunu-A}\\
  T^{(\rm SF)}_{\mu\nu}&=\nabla_\mu\phi\nabla_\nu\phi-g_{\mu\nu}\left(\frac{1}{2}(\nabla\phi)^2+\lambda\phi^4\right)\\&~~+\frac{1}{6}(g_{\mu\nu}\Box-\nabla_\mu\nabla_\nu+G_{\mu\nu})\phi^2\,.\notag
\end{align}
\end{subequations}
Since both Maxwell and Yang-Mills theory are invariant under conformal rescaling of the metric, $g_{\mu\nu}\mapsto \Omega^2g_{\mu\nu}$, with $\Omega=\Omega(x)$ an arbitrary local function, the corresponding stress--energy tensors $T_{\mu\nu}^{(\rm Max)}$ and $T_{\mu\nu}^{(\rm YM)}$ are traceless. On the other hand, the action for the scalar field~\eqref{I-f} is invariant (up to boundary terms) under simultaneous conformal transformations of the metric and the scalar field, $(g_{\mu\nu},\phi)\mapsto (\Omega^2g_{\mu\nu},\Omega^{-1}\phi)$; then $T_{\mu\nu}^{(\rm SF)}$ is traceless on-shell, since its trace is proportional to the scalar field equation. Thus, the spacetimes admitted by the theory have constant scalar curvature, that is, $R=4\Lambda$, which follows from taking the trace of Eq.~\eqref{EOM-g}.

\section{Hairy meronic black hole solution}\label{sec:meronic-black-hole-solution}
In order to seek static black hole solutions within the theory defined by the action~\eqref{action}, we adopt the following ansätze for the metric, the scalar field, and the Abelian gauge potential
\begin{subequations}\label{ansatz-ff} 
\begin{align}
\dd s^2&=-f(r)\dd t^2+\frac{\dd r^2}{f(r)}+r^2\dd\Omega^2_k\,,\\ \phi &=\phi(r)\,, \quad \A=-\frac{q}{r}\dd t + p\cos(\sqrt{k}\vartheta)\dd\varphi\,,
\end{align}
\end{subequations}
where $\dd\Omega_k^2$ denotes the line element of a two-dimensional Euclidean space of nonvanishing constant curvature $k \in \{-1,1\}$, corresponding to negative and positive horizon curvature, respectively, which can be parametrized by
\begin{equation}\label{dOmegak}
	\dd\Omega^2_k=\dd\vartheta^2+\frac{1}{k}\sin^2(\sqrt{k}\vartheta)\dd\varphi^2\,.
\end{equation}
It is straightforward to verify that with this choice of ansatz, the Maxwell equations~\eqref{eom-Maxwelll} are automatically satisfied. The integration constants $q$ and $p$ can be identified with the electric and magnetic charges of the solution defined as
\begin{equation}
    Q_e:=\frac{1}{4\pi }\int_{\Sigma_\infty}\star \F\quad \text{and}\quad Q_m:=-\frac{1}{4\pi }\int_{\Sigma_\infty}\F\,,
\end{equation}
respectively, where $\star$ denotes the Hodge dual, $\Sigma_\infty$ is the spatial infinity, and we have defined $\F=\frac{1}{2}\F_{\mu\nu}\dd x^\mu \wedge\dd x^\nu$ and $\star \F=\frac{1}{4}\epsilon_{\mu\nu\lambda\rho}\F^{\lambda\rho}\dd x^\mu\wedge \dd x^\nu$, where $\epsilon_{\mu\nu\lambda\rho}$ is the Levi-Civita tensor. For the ans\"atze in Eq.~\eqref{ansatz-ff} they read
\begin{equation}
	Q_e=\frac{\Sigma_k}{4\pi }q \qquad \text{and}\qquad  Q_m=\frac{\Sigma_k}{4\pi }p\,,
\end{equation}
where $\Sigma_k$ denotes the volume of the transverse section of constant curvature. 

To construct an ansatz for the meronic fields we closely follow the construction presented in Ref.~\cite{Canfora:2022nso}. There, the authors employ the so-called maximal embedding of $SU(2)$ into $SU(N)$, using an Euler-angle parametrization~\cite{Bertini:2005rc,Cacciatori:2012qi,Tilma:2004kp}, to build a basis of three $N\times N$ matrices that realizes an irreducible spin-$j$ representation of $SU(2)$, with $j=(N-1)/2$. Here, we extend this construction to include the noncompact group $SU(N-1,1)$. Therefore, we parametrize the generators of the non-Abelian gauge group as
\begin{subequations}\label{generators}
	\begin{align}
		t_1&=-\frac{i}{2}\sum_{j=2}^N\sqrt{(j-1)(N-j+1)}(E_{j-1,j}+E_{j,j-1})\,,\\
		t_2&=\frac{\sqrt{k}}{2}\sum_{j=2}^N\sqrt{(j-1)(N-j+1)}(E_{j-1,j}-E_{j,j-1})\,,\\
		t_3&=i\sqrt{k}\sum_{j=1}^N\left(\frac{N+1}{2}-j\right)E_{j,j}\,,
	\end{align}
\end{subequations}
where $(E_{i,j})_{mn}=\delta_{im}\delta_{jn}$, and are chosen such that they close the $\mathfrak{su}(N)$ algebra in the case of positive curvature ($k=1$), and the $\mathfrak{su}(N-1,1)$ algebra for negative curvature ($k=-1$), namely,
\begin{equation}
	[t_1,t_2]=t_3\,,\quad [t_2,t_3]=kt_1\,,\quad [t_3,t_1]=t_2\,,
\end{equation}
and are normalized according to
\begin{equation}
	\Tr(t_at_b)=-\frac{N(N^2-1)}{12}\diag(1,k,k)\,.
\end{equation}

Note that the choice of generators in Eq.~\eqref{generators} depends explicitly on the curvature of the black hole horizon. This feature is not new. In Ref.~\cite{Corral:2024xfv}, it was shown that New Massive Gravity~\cite{Bergshoeff:2009hq} admits anisotropic $SU(2)$ merons as solutions on squashed spheres~\cite{Canfora:2023bug}, which correspond to spaces of positive curvature. However, in order to construct analogous configurations on three-dimensional warped Anti-de Sitter (WAdS$_3$), a space of negative curvature, it becomes necessary to perform an analytic continuation of the gauge theory to the noncompact group $SU(1,1)$~\cite{Corral:2024xfv}. This provides further evidence that the internal gauge group of non-Abelian fields is not independent of the underlying topology, but rather that the two structures are intimately related. For planar horizons ($k=0$), the internal group reduces to an Abelian one, and therefore this case will not be considered in the remainder of this work. Recently, this strategy based on maximal embeddings of the gauge group has been widely employed to construct self-gravitating solutions in nonlinear field theories~\cite{Ayon-Beato:2019tvu,Barriga:2025mky,Vera:2025qqz,Henriquez-Baez:2024rjb,Concha:2023qcs,Henriquez-Baez:2022ubu,Flores-Alfonso:2024gag,Canfora:2022nso,Cacciatori:2021neu,Alvarez:2020zui,Cacciatori:2022kag}.

Having established this, we consider for the non-Abelian gauge potential a meron-type ansatz of the form~\cite{Canfora:2012ap}
\begin{equation}\label{A}
	A=\frac{1}{2e} U^{-1}\dd U\,,
\end{equation}
where the scalar $U(x)$ is an element of the non-Abelian group under consideration, and we choose the following parametrization
\begin{equation}\label{U}
	U=e^{-\varphi t_3}e^{2\vartheta t_2}e^{\varphi t_3}\,.
\end{equation}
Explicitly, the components of $A=A^{a}t_a$ along the group generators read
\begin{subequations}
	\begin{align}
		eA^1&=\sqrt{k}\sin(\sqrt{k}\varphi)\dd\vartheta+\frac{\sqrt{k}}{2}\cos(\sqrt{k}\varphi)\sin(2\sqrt{k}\vartheta)\dd\varphi\,,\\
		eA^2&=\cos(\sqrt{k}\varphi)\dd\vartheta-\frac{1}{2}\sin(\sqrt{k}\varphi)\sin(2\sqrt{k}\vartheta)\dd\varphi\,,\\
		eA^3&=\sin^2(\sqrt{k}\vartheta)\dd\varphi\,,
	\end{align}
\end{subequations}
while the components of the field strength $F=F^{a}t_a$ [cf.~Eq~\eqref{F-YM}] take the form
\begin{subequations}
	\begin{align}
		eF^1&=-k\cos(\sqrt{k}\varphi)\sin^2\left(\sqrt{k}\vartheta\right)\dd\vartheta\wedge\dd\varphi\,,\\
		eF^2&=\sqrt{k}\sin(\sqrt{k}\varphi)\sin^2\left(\sqrt{k}\vartheta\right)\dd\vartheta\wedge\dd\varphi\,,\\
		eF^3&=\sqrt{k}\sin(\sqrt{k}\vartheta)\cos(\sqrt{k}\vartheta)\dd\vartheta\wedge\dd\varphi\,,
	\end{align}
\end{subequations}
from which it is clear that these configurations are genuinely non-Abelian. It is straightforward to verify that this ansatz identically satisfies the Yang-Mills equations~\eqref{EOM-A}, while the remaining field equations are solved by
\begin{subequations}
	\begin{align}
		f(r)&=k\left(1-\frac{M G}{r}\right)^2-\frac{\Lambda}{3}r^2\,,\\
		\phi(r)&=\sqrt{-\frac{\Lambda}{6\lambda}}\frac{M G}{(r-MG)}\label{sol-phi}\,,
	\end{align}
\end{subequations}
where $M$ is an integration constant related with the mass of the solution. Together with the remaining integration constants arising from the Maxwell fields, it is subject to the constraint
\begin{equation}\label{constraint}
	\frac{q^2+ p^2}{k}=\frac{M^2G}{4\pi}\left(1+\frac{2\pi \Lambda G}{9\lambda}\right	)-\frac{N(N^2-1)}{6e^2}\,.
\end{equation}
This solution generalizes the MTZ black hole~\cite{Martinez:2002ru,Martinez:2005di} to the presence non-Abelian gauge fields. The existence of an event horizon requires the sign of the curvature of the base manifold and that of the cosmological constant to coincide. However, the requirement that the scalar field be real forces the cosmological constant and the conformal potential coupling to have opposite signs. This, in turn, imposes additional constraints on the parameters, which are discussed below. 

For horizons of positive curvature, $k=1$, the solution describes a black hole provided that the cosmological constant is positive, namely $\Lambda=3\ell^{-2}$, where $\ell$ denotes the de Sitter radius, and
\begin{equation}\label{lambda-positivo}
	\lambda<\frac{2\pi(MGe)^2}{\ell^2[2\pi N(N^2-1)-3M^2e^2G]}<0\,.
\end{equation}
For this case, the geometry corresponds to an extremal black hole enclosed by a cosmological horizon, a configuration known as a lukewarm black hole.

The inner horizon, the event horizon, and the cosmological horizon are located at
\begin{subequations}\label{zeros-positivo}
	\begin{align}
		r_-&=\frac{\ell}{2}\left(-1+\sqrt{1+\frac{4MG}{\ell}}\right)\,,\\
		r_+&=\frac{\ell}{2}\left(1-\sqrt{1-\frac{4MG}{\ell}}\right)\,,\\
		r_{++}&=\frac{\ell}{2}\left(1+\sqrt{1-\frac{4MG}{\ell}}\right)\,,
	\end{align}
\end{subequations}
respectively. 
Moreover, the mass is bounded both from below and from above, namely
\begin{equation}\label{Mmin}
	\frac{1}{e}\sqrt{\frac{2\pi N(N^2-1)}{3G}}<M<\frac{\ell}{4 G}\,.
\end{equation}
It is clear that the massless limit is not continuously connected whenever the meronic configuration is present. When $M=\ell/4G$, the event and cosmological horizons coincide, and the solution reduces to the Nariai geometry.

On the other hand, when the horizon has negative curvature, $k=-1$, the cosmological constant must also be negative, $\Lambda=-3\ell^{-2}$, in order to guarantee the existence of real zeros of the lapse function. In this case, requiring the scalar field to be well defined imposes that the coupling of the conformal potential satisfies
\begin{equation}
	0<\lambda <\frac{2\pi (MGe)^2}{\ell^2[3M^2e^2G-2\pi N(N^2-1)]}\,.
\end{equation}
In this case, the mass is also bounded from below as in the previous case, and the geometry corresponds to a black hole with multiple horizons.

On the other hand, the scalar field develops a singularity at $r=MG$, which lies behind the event horizon, in contrast to the BBMB black hole~\cite{Bocharova:1970,Bekenstein:1974sf}. The causal structure, thermodynamic properties, and stability of the MTZ black hole have been studied in Refs.~\cite{Barlow:2005yd,Harper:2003wt}. We will not discuss this geometry in further detail, as it has been extensively analyzed in the literature; see, for instance, Refs.~\cite{Salgado:2003ub,Ashtekar:2003jh,Ashtekar:2003zx,McFadden:2004ni,Salgado:2005hx}.

In the limit of vanishing cosmological constant, $\Lambda \to 0$, which implies $\lambda = 0$ due to the constraint in Eq.~\eqref{constraint}, the solution reduces to
\begin{widetext}
 \begin{subequations}
 	\begin{align}
 		\dd s^2&=-k\left(1-\frac{MG}{r}\right)^2\dd t^2+\left[k\left(1-\frac{MG}{r}\right)^2\right]^{-1}\dd r^2 + r^2\left(\dd \vartheta^2+\frac{1}{k}\sin^2(\sqrt{k}\vartheta)\dd\varphi^2\right)\,,\\
 	\phi(r)&=\frac{1}{2e}\sqrt{\frac{3kGe^2M^2-2\pi k N(N^2-1)-12\pi^2e^2\left(p^2+q^2\right)}{\pi k}}\frac{1}{(r-MG)}\,,
 	\end{align}
 \end{subequations}
\end{widetext}
where the gauge potential is still given by Eq.~\eqref{A}, and the mass bound remains determined by Eq.~\eqref{Mmin}. This configuration represents a generalization of the dyonic BBMB black hole~\cite{Bekenstein:1974sf,Bocharova:1970}, sourced by self-gravitating meronic fields, and exhibits a scalar-field singularity located at the horizon $r=MG$. Moreover, the scalar field becomes trivial when $M=M_{\min}$ and the Maxwell field vanishes, in which case the extremal meronic black hole is recovered~\cite{Canfora:2022nso}.

\section{(Super-)renormalizable meronic spacetimes}\label{sec:super-renormalizable}
In the previous section, the dyonic MTZ black hole was generalized to include self-gravitating Yang-Mills fields. In this section, we construct a new solution by employing the strategy proposed in Ref.~\cite{Ayon-Beato:2015ada}. There, the authors devised a method to generate solutions by mapping any self-gravitating conformal scalar field with non-trivial self-interaction and a nonvanishing cosmological constant into a one-parameter family of a related theory, supplemented by a (super-)renormalizable potential. When the MTZ black hole is used as the seed, the resulting geometry is nothing but the solution previously reported by Anabal\'on and Cisterna (AC) in Ref.~\cite{Anabalon:2012tu}. This family of solutions includes regular black holes, wormholes, and bouncing cosmologies, thereby significantly enhancing the solution space compared to the original seed. This strategy was extended to incorporate Maxwell fields as well~\cite{Ayon-Beato:2015ada}, which is possible thanks to their conformal invariance. 
This technique has also been implemented using geometries with NUT charge~\cite{Bardoux:2013swa} as seed in Ref.~\cite{Barrientos:2022avi}. Here, we apply a similar procedure to generate a new solution using the meronic MTZ black hole obtained in the previous Section as the conformal seed.

 The idea is to map the fields of the theory, together with the coupling constants, in such a way to obtain a new action that is proportional to the original one, namely,
\begin{equation}\label{Ibar-I}
	\bar{I}[\bar{g}_{\mu \nu},\bar{\A}_\mu,\bar{A}_\mu,\bar{\phi}]=(1-a^2)I[g_{\mu\nu},\A_\mu,A_\mu,\phi]\,,
\end{equation}
where $a$ is the parameter that controls the mapping and is restricted by $a^2<1$ in order to preserve unitarity in both theories.
As a consequence, the solutions of the equations of motion~\eqref{EOM}, obtained by performing stationary variations of the right-hand side of Eq.~\eqref{Ibar-I}, can be mapped directly onto the field equations arising from varying the left-hand side. 

The field equations of the new theory are given by
\begin{subequations}
	\begin{align}
  \bar{G}_{\mu\nu}+\bar{\Lambda}\bar{g}_{\mu\nu}&=8\pi G\left(\bar{T}_{\mu\nu}^{\rm (Max)}+\bar{T}_{\mu\nu}^{(\rm YM)} + \bar{T}_{\mu\nu}^{(\rm SF)}\right)\,,\\
 \bar{\nabla}_\mu\bar{\F}^{\mu\nu}&=0\,,\\
 \bar{\nabla}_\mu\bar{F}^{\mu\nu}+\bar{e}[\bar{A}_\mu,\bar{F}^{\mu\nu}]&=0\,,\\
  \bar{\Box}\bar{\phi}&=\frac{1}{6}\bar{R}\bar{\phi}+\dv{V(\bar{\phi})}{\bar{\phi}}\,,
\end{align}
\end{subequations}
where the barred stress-energy tensor associated with the scalar field is given by
\begin{equation}
\begin{split}
    \bar{T}_{\mu\nu}^{(\rm SF)}&=\bar{\nabla}_\mu\bar{\phi}\bar{\nabla}_\nu\bar{\phi}-\bar{g}_{\mu\nu}\left(\frac{1}{2}(\bar{\nabla}\bar{\phi})^2+V(\bar{\phi})\right)\\
    &~~+\frac{1}{6}\left(\bar{g}_{\mu\nu}\bar{\Box}-\bar{\nabla}_\mu\bar{\nabla}_\nu+\bar{G}_{\mu\nu}\right)\bar{\phi}^2\,,
\end{split}
\end{equation}
whereas the contribution associated with the gauge fields takes the same form as that defined in Eq.~\eqref{Tmunu-A}, but is constructed from the barred quantities.  Moreover,
\begin{subequations}
\begin{align}
	\bar{\F}_{\mu\nu}&=\partial_\mu\bar{\A}_\nu-\partial_\nu\bar{\A}_\mu\,,\\
    \bar{F}_{\mu\nu}&=\partial_\mu \bar{A}_\nu -\partial_\nu \bar{A}_\mu +\bar{e}[\bar{A}_\mu,\bar{A}_\nu ]\,,
\end{align}
\end{subequations}
denotes the Maxwell and Yang-Mills field strength of the new theory, respectively, while the (super-)renormalizable scalar potential is given by

\begin{equation}
	V(\bar{\phi})=\alpha_1\bar{\phi}+\alpha_2\bar{\phi}^2+\alpha_3\bar{\phi}^3+\alpha_4\bar{\phi}^4\,,
\end{equation}
where the coupling constants take the following form
\begin{subequations}
	\begin{align}
  \alpha_1&=\frac{a\sqrt{3}(2\Lambda G\pi+9a^2\lambda)}{6(a^2-1)^3(\pi G)^\frac{3}{2}}\,,\\
  \alpha_2&=-\frac{a^2\left(2\Lambda G\pi +9\lambda\right)}{2(a^2-1)^3\pi G}\,,\\
  \alpha_3&=\frac{2a\sqrt{3}(2\Lambda G\pi a^2+9\lambda)}{9(a^2-1)^3\sqrt{\pi G}}\,,\\
  \alpha_4&=-\frac{2\Lambda \pi G a^4+9\lambda}{9(a^2-1)^3}\,.
\end{align}
\end{subequations}
The cosmological constant and the Yang-Mills coupling are given by
\begin{subequations}
\begin{align}
	\bar{\Lambda}&=-\frac{9a^4\lambda+2\pi \Lambda G}{2\pi G(a^2-1)^3} \,,\\
    \bar{e}&=\frac{e}{\sqrt{1-a^2}}\,,
\end{align}
\end{subequations}
respectively. Under this mapping, the meronic MTZ solution is mapped into a new variety of exact solutions sourced by non-Abelian gauge fields, given by
\begin{widetext}
	\begin{subequations}\label{meronic-AC}
		\begin{align}
	  \dd\bar{s}^2&= \left(1+\frac{a M\sqrt{-\frac{2\Lambda\pi G^3}{\lambda}}}{3(M G-r)}\right)^2 \Bigg\{-\left[k\left(1-\frac{M G}{r}\right)^2-\frac{\Lambda r^2}{3}\right]\dd t^2+\left[k\left(1-\frac{M G}{r}\right)^2-\frac{\Lambda r^2}{3}\right]^{-1}\dd r^2+r^2\dd\Omega^2_{k}\Bigg\}\,,\\
  \bar{\phi}(r)&=\frac{1}{2\sqrt{G}}\frac{3a\sqrt{\frac{3}{\pi}}(MG-r)+M\sqrt{-\frac{6\Lambda G^3}{\lambda}}}{3(MG-r)+aM\sqrt{-\frac{2\pi\Lambda G^3}{\lambda}}} \,,\qquad \bar{\A}=\sqrt{1-a^2}\left[-\frac{q}{r}\dd t + p\cos(\sqrt{k}\vartheta)\dd\varphi\right]\,,\qquad  \bar{A}=\frac{1}{2\bar{e}}U^{-1}\dd U\,,
\end{align}

	\end{subequations}
\end{widetext}
where $U$ is given by Eq.~\eqref{U}, $\dd\Omega^2_k$ is defined in Eq.~\eqref{dOmegak}, and the coupling constants satisfy the constraint in Eq.~\eqref{constraint}. 

These geometries correspond to the dyonic generalization of the AC solution, supported by meronic fields and a (super-)renormalizable potential. The presence of both Abelian and non-Abelian gauge fields allows for the inclusion of a mass term in the scalar potential, thus enabling the complete series of super-renormalizable terms in the scalar self-interaction potential. The causal structure of the geometry described by Eq.~\eqref{meronic-AC} is significantly enhanced compared to that of the meronic MTZ solution, giving rise to configurations ranging from black holes to regular inhomogeneous bouncing cosmologies and wormholes.  These cases are discussed in detail in Ref.~\cite{Anabalon:2012tu}.

The electric and magnetic charges are shifted due to the presence of the parameter $a$. In the limit of vanishing Abelian fields, $q=p=0$, the solution becomes rigid, with the mass constrained by the couplings of the theory through the condition~\eqref{constraint}. Finally, we note that for $a=0$ the mapping reduces to the identity, and the meronic MTZ black hole reported in the previous section is recovered. However, new special geometries arise as particular limits obtained by switching off parameters in the super-renormalizable solution~\eqref{meronic-AC}, which we do not discuss in detail, as they are straightforward to obtain.

\section{Non-Noetherian conformal meronic black hole}\label{sec:nN-meronic-BH}
Recently, in Ref.~\cite{Ayon-Beato:2023bzp}, the most general action for a real scalar field leading to a conformally invariant second-order equation of motion was constructed. This corresponds to a generalization of the theory presented in Ref.~\cite{Fernandes:2021dsb}. The associated scalar field equation includes a new nonminimal coupling that breaks the conformal invariance of the action, while preserving it at the level of the equations of motion. Such a symmetry was dubbed non-Noetherian conformal symmetry~\cite{Ayon-Beato:2023bzp}, since it is not realized at the level of the action principle (see also Refs.~\cite{Jackiw:2005su,Ayon-Beato:2023lrn} for the analogous construction in lower dimensions). This theory has given rise to several solutions of physical interest, such as black holes~\cite{Fernandes:2021dsb,Ayon-Beato:2024vph,Babichev:2022awg}, cosmological models~\cite{Fernandes:2021dsb}, stealth configurations~\cite{Ayon-Beato:2024xgp}, and exact wave solutions~\cite{Sanchez:2024xke}, among others. In addition, its DC conductivity properties have been explored in Ref.~\cite{Hernandez-Vera:2024zui} (see Ref.~\cite{Fernandes:2022zrq} for a comprehensive review).

In this Section we consider an extension of the theory presented in Sec.~\ref{sec:the-theory}, in which the conformal non-Noetherian sector associated with the scalar field is taken into account. The action principle to be considered is given by
\begin{equation}
    I[g_{\mu\nu},\A_\mu ,A_\mu,\phi]=I_{\rm G}+I_{\rm Max} + I_{\rm YM}+I_{\rm SF}+I_{\rm nN}\,,
\end{equation}
where the first four terms are defined in Eq.~\eqref{I's}, and the non-Noetherian contribution is given by~\cite{Fernandes:2021dsb,Ayon-Beato:2023bzp}
    \begin{align}\label{I+nN}
        I_{\rm nN}&=-\frac{\alpha}{2}\int_\mathcal{M}\dd^4x\sqrt{|g|}\left(\log(\phi)\mathcal{G}-\frac{4}{\phi^2}G^{\mu\nu}\nabla_\mu\phi\nabla_\nu\phi\right.\notag\\
        &\left.~~-\frac{4}{\phi^3}(\nabla\phi)^2\Box\phi+\frac{2}{\phi^4}(\nabla\phi)^4\right)\,,
    \end{align}
where $\mathcal{G}=R^2-4R_{\mu\nu}R^{\mu\nu}+R_{\mu\nu\lambda\rho}R^{\mu\nu\lambda\rho}$ is the Gauss-Bonnet term. This additional contribution to the action belongs to the Horndeski class~\cite{Horndeski:1974wa,Babichev:2023dhs,Babichev:2023rhn} and is closely related to formulations of the so-called $4D$-Einstein-Gauss-Bonnet class of theories~\cite{Glavan:2019inb,Lu:2020iav,Fernandes:2020nbq,Hennigar:2020lsl,Fernandes:2022zrq}.

The field equation associated with the scalar field is corrected with respect to Eq.~\eqref{EOM-f} by the addition of a new conformally invariant term, namely,
\begin{equation}\label{EOM-f-nN}
    \Box\phi-\frac{1}{6}R\phi-4\lambda\phi^3-\left(4\lambda+\frac{\alpha}{2}\tilde{\mathcal{G}}\right)\phi^3=0\,,
\end{equation}
where $\tilde{\mathcal{G}}$ denotes the Gauss-Bonnet term constructed from the auxiliary metric
\begin{equation}\label{g-tilde}
    \tilde{g}_{\mu\nu}=\phi^2 g_{\mu\nu}\,,
\end{equation}
which is manifestly conformally invariant under the transformations discussed in Sec.~\ref{sec:the-theory}. The contribution to the energy-momentum tensor arising from the non-Noetherian scalar sector is presented explicitly in Appendix.~\ref{app:nN-Tmunu}. It is worth noting that the most general non-Noetherian action for a conformally coupled scalar field also includes an arbitrary function constructed from the auxiliary metric~\eqref{g-tilde} and its associated Weyl tensor. However, the inclusion of such a term leads to higher-derivative contributions in the stress-energy tensor due to the explicit presence of the Weyl tensor~\cite{Ayon-Beato:2023bzp}. By contrast, the extension in Eq.~\eqref{I+nN} yields corrections to the stress-energy tensor that remain second order in derivatives, since the corresponding Lagrangian belongs to the Horndeski class~\cite{Horndeski:1974wa}. For this reason, in this Section we do not consider the most general arbitrary contributions that would spoil the second-order nature of the field equations.

In order to seek for static black hole solutions, we adopt the ans\"atze in Eq.~\eqref{ansatz-ff}, along with Eqs.~\eqref{A} and~\eqref{U}. To integrate de system, first we take the trace of the Einstein equations which yields a second-order partial differential equation for the metric function $f(r)$ which is solved by
\begin{widetext}
\begin{equation}\label{fsol-nN}
        f^{(i)}(r)=k+\frac{r^2}{16\alpha\pi G}\left[1\pm \sqrt{1+32\alpha\pi G\left(\frac{\Lambda}{3}+\frac{2MG}{r^3}-\frac{\mu^{(i)}}{r^4}\right)}\right]\,,
\end{equation}
\end{widetext}
where $M$ is an integration constant related with the mass of the solution, and $\mu^{(i)}$ is a constant to be fixed by the remaining equations.  We now observe that subtracting the $tt$ and $rr$ components of the Einstein equations yields an expression that can be factorized as
\begin{equation}\label{tt-rr}
    \left(\frac{\phi'}{\phi^2}\right)'\left[f\phi'(r^2\phi)'+(f-k\alpha)\phi^2-\frac{1}{12} r^2\phi^4\right]=0\,,
\end{equation}
which admits two distinct branches of non-trivial solutions for the scalar field. The first one corresponds to solving the first parentheses in the equation above, yielding
\begin{subequations}\label{phi-1-branch}
    \begin{align}
    \phi^{(1)}(r)&=\frac{2\sqrt{-3\alpha k}}{r}\label{phi-1}\,,\\
     \mu^{(1)}&=4\pi G \left(\frac{N(N^2-1)k}{6e^2}+p^2+q^2-4\alpha\right)\,,\\
    \lambda\alpha&=\frac{1}{144}\,,
    \end{align}
\end{subequations}
while the second branch arises from solving the second parenthesis in Eq.~\eqref{tt-rr}, which is solved by
\begin{subequations}\label{phi-2-branch}
    \begin{align}
       \phi^{(2)}(r) &=
\begin{cases}
\dfrac{2\sqrt{3 k \alpha}}{\,r\,\sinh\!\left(\sqrt{k}\left[\nu \pm \displaystyle\int^r \frac{\dd \rho}{\rho\sqrt{f^{(2)}(\rho)}}\right]\right)}\,, 
& \alpha > 0\,, \\[4.0em]
\dfrac{2\sqrt{-3 k \alpha}}{\,r\,\cosh\!\left(\sqrt{k}\left[\nu \pm \displaystyle\int^r \frac{\dd \rho}{\rho\sqrt{f^{(2)}(\rho)}}\right]\right)}\,, 
& \alpha < 0\,,
\end{cases}\\
 \mu^{(2)}&=\frac{2N(N^2-1)\pi G k}{3e^2}+4\pi G\left(q^2+p^2\right)\,,\\
    \lambda\alpha&=\frac{1}{48}\,.
    \end{align}
\end{subequations}
where $\nu$ is an unconstrained integration constant and may therefore be interpreted as an independent scalar hair parameter. Determining whether this constant contributes independently to the global charges requires a dedicated analysis of conserved charges in the non-Noetherian framework. We leave this issue for future work.

Focusing on the metric function $f(r)$ in Eq.~\eqref{fsol-nN}, we observe that in the limit where the non-Noetherian coupling vanishes, $\alpha\to 0$, the negative branch smoothly reduces to the static solution of the Einstein-Maxwell system with a cosmological constant
\begin{equation}
    f^{(i)}_-(r)=k-\frac{2MG}{r}-\frac{\Lambda r^2}{3}+\frac{\mu^{(i)}}{r^2} + \mathcal{O}(\alpha)\,,
\end{equation}
whereas the positive branch behaves as
\begin{equation}
    f^{(i)}_+(r)=\frac{r^2}{8\pi G\alpha}+k+\frac{2MG}{r}+\frac{\Lambda r^2}{3}-\frac{\mu^{(i)}}{r^2}+\mathcal{O}(\alpha)\,.
\end{equation}
The latter is typically discarded as unphysical, as it lacks a smooth $\alpha\to 0$ limit and displays the incorrect sign for the mass term. The thermodynamic analysis of this geometry has been carried out in Refs.~\cite{Fernandes:2021dsb,Fernandes:2022zrq,Fernandes:2021ysi,Fernandes:2020rpa}.

As for the first branch in Eq.~\eqref{phi-1-branch}, it is continuously connected to the solution reported in Ref.~\cite{Fernandes:2021dsb} once the non-Abelian sector of the theory is switched off and $k=1$ is assumed. On the other hand, the second branch in Eq.~\eqref{phi-2-branch} corresponds to a class of solutions that go beyond the standard conformally coupled scalar field theory, which is consistent with recent results on more general geometries admitted within Horndeski theory (see, for instance, Refs.~\cite{Babichev:2023rhn,Hassaine:2023paj}).

\section{Discussion}\label{sec:discussion}
In this work, we have constructed a broad class of exact solutions within Einstein--Maxwell--Yang-Mills theory with cosmological constant, and non-minimally coupled scalar fields. The gauge configurations considered correspond to merons, reflecting the intrinsically non-Abelian nature of these solutions. Moreover, the choice of gauge group depends explicitly on the topology of the horizon under consideration. For horizons of positive curvature, the group is $SU(N)$, whereas for negative curvature it is $SU(N-1,1)$. To parametrize the generators of these groups, we employ the maximal embedding of $SU(2)$ into $SU(N)$ [or $SU(1,1)$ into $SU(N-1,1)$], together with the Euler angle parametrization~\cite{Bertini:2005rc,Cacciatori:2012qi,Tilma:2004kp}, which allows us to construct solutions for arbitrary values of $N$.

We begin by considering the case of a conformally coupled scalar field, presenting the non-Abelian extension of the MTZ black hole~\cite{Martinez:2002ru,Martinez:2005di}. Subsequently, using this configuration as a conformal seed and applying an extension of the mappings proposed in Ref.~\cite{Ayon-Beato:2015ada}, we obtain a generalization of the AC~\cite{Anabalon:2012tu} solution in the presence of self-gravitating merons. Depending on the values of the coupling constants, this geometry admits particular limits describing black holes, inhomogeneous bouncing cosmologies, and wormholes. Finally, we analyze the non-Noetherian conformal extension of scalar fields~\cite{Fernandes:2021dsb,Ayon-Beato:2023bzp}, which breaks conformal invariance at the level of the action while preserving it in the scalar field equation. We show that the black hole solutions presented in Ref.~\cite{Fernandes:2021dsb} can be consistently extended to include self-gravitating Yang-Mills fields.

Several generalizations of the results presented in this work are conceivable. First of all, it is natural to ask whether our construction can be extended to higher dimensions, for instance to $D=5$, where a self-gravitating meronic black hole solution is known to exist and carries a nontrivial topological charge~\cite{Canfora:2018ppu}, in contrast to the solutions discussed in this work. It would also be interesting to explore the possibility of constructing self-gravitating Yang-Mills configurations on spacetimes endowed with NUT charge and self-gravitating scalar conformal scalar fields, as in, e.g., Ref.~\cite{Bardoux:2013swa}. If feasible, such configurations would provide a one-parameter extensions of the solutions obtained here, allowing for a more detailed study of their thermodynamic and holographic properties. Extensions incorporating the recently proposed non-Abelian generalizations of ModMax electrodynamics~\cite{Bandos:2020jsw,Canfora:2025gwm,Cirilo-Lombardo:2023poc}, which preserve conformal invariance, could also be of interest,  and we expect that they reduce to the solutions presented in this work in the limit where the dimensionless non-Abelian ModMax parameter vanishes. Furthermore, exploring the compatibility of meronic configurations with geometries with nontrivial torsion (see, for example, Ref.~\cite{Aviles:2024muk}) would undoubtedly constitute an interesting extension of these results.  Finally, it is worth emphasizing that since the non-Noetherian sector lacks conformal invariance at the level of the action, it is not clear whether a mapping analogous to that discussed in Sec.~\ref{sec:super-renormalizable} can be implemented when this sector is taken into account which remains an open problem that is certainly worth exploring. We leave these questions for future work.

\begin{acknowledgments}
    The authors thank Eloy Ay\'on-Beato, Fabrizio Canfora, Crist\'obal Corral, Oscar Fuentealba, Ulises Hern\'andez-Vera and Marcela Lagos for insightful comments and discussions. The work of L.A. is partially supported by Agencia Nacional de Investigaci\'{o}n y Desarrollo (ANID) through Fondecyt Iniciación Grant No.~11261098 and SIA-ANID Grant No.~85220027. The authors would also like to thank the NordGrav 2026 conference, held in Iquique, Chile, for providing a stimulating and welcoming environment where a significant part of the discussions leading to this work took place.
\end{acknowledgments}

\appendix

\section{Non-Noetherian energy-momentum tensor contribution}\label{app:nN-Tmunu}
In the presence of the non-Noetherian conformal contribution of the scalar field~\eqref{I+nN}, the Einstein field equations take the form
\begin{equation}
    G_{\mu\nu}+\Lambda g_{\mu\nu}=8\pi G\left(T^{(\rm Max)}_{\mu\nu}+T_{\mu\nu}^{(\rm YM)}+T^{(\rm SF)}_{\mu\nu}+\alpha T^{\rm nN}_{\mu\nu}\right)\,,
\end{equation}
where the first contributions to the energy-momentum tensor are given in Eq.~\eqref{Tmunu}, while the non-Noetherian contribution is given by
\begin{widetext}
    \begin{equation}
        \begin{split}
            T_{\mu\nu}^{\rm nN}&=-\frac{2G_{\mu\nu}(\nabla\phi)^2}{\phi^2}+2P_{\mu\alpha\nu\beta}\left(\frac{2\phi^\alpha\phi^\beta}{\phi^2}-\frac{\phi^{\alpha\beta}}{\phi}\right)-4\left(\frac{2\phi_\alpha\phi_\mu}{\phi^2}-\frac{\phi_{\alpha\mu}}{\phi}\right)\left(\frac{2\phi^\alpha\phi_\nu}{\phi^2}-\frac{\phi^\alpha_{~\nu}}{\phi}\right)\\
            &~~~-4\left(\frac{2\phi_\mu\phi_\nu}{\phi^2}-\frac{\phi_{\mu\nu}}{\phi}\right)\left(\frac{\Box\phi}{\phi}-\frac{(\nabla\phi)^2}{\phi^2}\right)-
       g_{\mu\nu}\left[2\left(\frac{\Box\phi}{\phi}-\frac{(\nabla\phi)^2}{\phi^2}\right)^2-\frac{(\nabla\phi)^4}{\phi^4}\right]\\
       &~~~-2g_{\mu\nu}\left(\frac{\phi_{\alpha\beta}}{\phi}-\frac{\phi_\alpha\phi_\beta}{\phi^2}\right)\left(\frac{3\phi^\alpha\phi^\beta}{\phi^2}-\frac{\phi^{\alpha\beta}}{\phi}\right)\,.
        \end{split}
    \end{equation}
\end{widetext}
Here, we have used the convention that indices acting on scalar quantities denote covariant derivatives, and we have further defined
\begin{equation}
    P^{\mu\nu}_{\alpha\beta}=\pdv{\mathcal{G}}{R_{\mu\nu}^{\alpha\beta}}=-\frac{1}{2}\delta^{\mu\nu\lambda\rho}_{\alpha\beta\tau\sigma}R^{\tau\sigma}_{\lambda\rho}\,,
\end{equation}
as the variation of the Gauss-Bonnet term with respect to the Riemann tensor. The field equations for the Yang-Mills and the scalar fields are given by Eqs.~\eqref{EOM-f} and~\eqref{EOM-f-nN}, respectively.

\bibliography{References}

\begin{thebibliography}{125}%
\makeatletter
\providecommand \@ifxundefined [1]{%
 \@ifx{#1\undefined}
}%
\providecommand \@ifnum [1]{%
 \ifnum #1\expandafter \@firstoftwo
 \else \expandafter \@secondoftwo
 \fi
}%
\providecommand \@ifx [1]{%
 \ifx #1\expandafter \@firstoftwo
 \else \expandafter \@secondoftwo
 \fi
}%
\providecommand \natexlab [1]{#1}%
\providecommand \enquote  [1]{``#1''}%
\providecommand \bibnamefont  [1]{#1}%
\providecommand \bibfnamefont [1]{#1}%
\providecommand \citenamefont [1]{#1}%
\providecommand \href@noop [0]{\@secondoftwo}%
\providecommand \href [0]{\begingroup \@sanitize@url \@href}%
\providecommand \@href[1]{\@@startlink{#1}\@@href}%
\providecommand \@@href[1]{\endgroup#1\@@endlink}%
\providecommand \@sanitize@url [0]{\catcode `\\12\catcode `\$12\catcode `\&12\catcode `\#12\catcode `\^12\catcode `\_12\catcode `\%12\relax}%
\providecommand \@@startlink[1]{}%
\providecommand \@@endlink[0]{}%
\providecommand \url  [0]{\begingroup\@sanitize@url \@url }%
\providecommand \@url [1]{\endgroup\@href {#1}{\urlprefix }}%
\providecommand \urlprefix  [0]{URL }%
\providecommand \Eprint [0]{\href }%
\providecommand \doibase [0]{http://dx.doi.org/}%
\providecommand \selectlanguage [0]{\@gobble}%
\providecommand \bibinfo  [0]{\@secondoftwo}%
\providecommand \bibfield  [0]{\@secondoftwo}%
\providecommand \translation [1]{[#1]}%
\providecommand \BibitemOpen [0]{}%
\providecommand \bibitemStop [0]{}%
\providecommand \bibitemNoStop [0]{.\EOS\space}%
\providecommand \EOS [0]{\spacefactor3000\relax}%
\providecommand \BibitemShut  [1]{\csname bibitem#1\endcsname}%
\let\auto@bib@innerbib\@empty
\bibitem [{\citenamefont {Bizon}(1990)}]{Bizon:1990sr}%
  \BibitemOpen
  \bibfield  {author} {\bibinfo {author} {\bibfnamefont {P.}~\bibnamefont {Bizon}},\ }\href {\doibase 10.1103/PhysRevLett.64.2844} {\bibfield  {journal} {\bibinfo  {journal} {Phys. Rev. Lett.}\ }\textbf {\bibinfo {volume} {64}},\ \bibinfo {pages} {2844} (\bibinfo {year} {1990})}\BibitemShut {NoStop}%
\bibitem [{\citenamefont {Smoller}\ and\ \citenamefont {Wasserman}(1997)}]{Smoller:1997qr}%
  \BibitemOpen
  \bibfield  {author} {\bibinfo {author} {\bibfnamefont {J.~A.}\ \bibnamefont {Smoller}}\ and\ \bibinfo {author} {\bibfnamefont {A.~G.}\ \bibnamefont {Wasserman}},\ }\href {\doibase 10.1063/1.532224} {\bibfield  {journal} {\bibinfo  {journal} {J. Math. Phys.}\ }\textbf {\bibinfo {volume} {38}},\ \bibinfo {pages} {6522} (\bibinfo {year} {1997})},\ \Eprint {http://arxiv.org/abs/gr-qc/9703062} {arXiv:gr-qc/9703062} \BibitemShut {NoStop}%
\bibitem [{\citenamefont {Winstanley}(2009)}]{Winstanley:2008ac}%
  \BibitemOpen
  \bibfield  {author} {\bibinfo {author} {\bibfnamefont {E.}~\bibnamefont {Winstanley}},\ }\href {\doibase 10.1007/978-3-540-88460-6_2} {\bibfield  {journal} {\bibinfo  {journal} {Lect. Notes Phys.}\ }\textbf {\bibinfo {volume} {769}},\ \bibinfo {pages} {49} (\bibinfo {year} {2009})},\ \Eprint {http://arxiv.org/abs/0801.0527} {arXiv:0801.0527 [gr-qc]} \BibitemShut {NoStop}%
\bibitem [{\citenamefont {Volkov}\ and\ \citenamefont {Gal'tsov}(1999)}]{Volkov:1998cc}%
  \BibitemOpen
  \bibfield  {author} {\bibinfo {author} {\bibfnamefont {M.~S.}\ \bibnamefont {Volkov}}\ and\ \bibinfo {author} {\bibfnamefont {D.~V.}\ \bibnamefont {Gal'tsov}},\ }\href {\doibase 10.1016/S0370-1573(99)00010-1} {\bibfield  {journal} {\bibinfo  {journal} {Phys. Rept.}\ }\textbf {\bibinfo {volume} {319}},\ \bibinfo {pages} {1} (\bibinfo {year} {1999})},\ \Eprint {http://arxiv.org/abs/hep-th/9810070} {arXiv:hep-th/9810070} \BibitemShut {NoStop}%
\bibitem [{\citenamefont {Bartnik}\ and\ \citenamefont {Mckinnon}(1988)}]{Bartnik:1988am}%
  \BibitemOpen
  \bibfield  {author} {\bibinfo {author} {\bibfnamefont {R.}~\bibnamefont {Bartnik}}\ and\ \bibinfo {author} {\bibfnamefont {J.}~\bibnamefont {Mckinnon}},\ }\href {\doibase 10.1103/PhysRevLett.61.141} {\bibfield  {journal} {\bibinfo  {journal} {Phys. Rev. Lett.}\ }\textbf {\bibinfo {volume} {61}},\ \bibinfo {pages} {141} (\bibinfo {year} {1988})}\BibitemShut {NoStop}%
\bibitem [{\citenamefont {Shepherd}\ and\ \citenamefont {Winstanley}(2016)}]{Shepherd:2015dse}%
  \BibitemOpen
  \bibfield  {author} {\bibinfo {author} {\bibfnamefont {B.~L.}\ \bibnamefont {Shepherd}}\ and\ \bibinfo {author} {\bibfnamefont {E.}~\bibnamefont {Winstanley}},\ }\href {\doibase 10.1103/PhysRevD.93.064064} {\bibfield  {journal} {\bibinfo  {journal} {Phys. Rev. D}\ }\textbf {\bibinfo {volume} {93}},\ \bibinfo {pages} {064064} (\bibinfo {year} {2016})},\ \Eprint {http://arxiv.org/abs/1512.03010} {arXiv:1512.03010 [gr-qc]} \BibitemShut {NoStop}%
\bibitem [{\citenamefont {Shepherd}\ and\ \citenamefont {Winstanley}(2017)}]{Shepherd:2016ily}%
  \BibitemOpen
  \bibfield  {author} {\bibinfo {author} {\bibfnamefont {B.~L.}\ \bibnamefont {Shepherd}}\ and\ \bibinfo {author} {\bibfnamefont {E.}~\bibnamefont {Winstanley}},\ }\href {\doibase 10.1007/JHEP01(2017)065} {\bibfield  {journal} {\bibinfo  {journal} {JHEP}\ }\textbf {\bibinfo {volume} {01}},\ \bibinfo {pages} {065} (\bibinfo {year} {2017})},\ \Eprint {http://arxiv.org/abs/1611.04162} {arXiv:1611.04162 [hep-th]} \BibitemShut {NoStop}%
\bibitem [{\citenamefont {de~Alfaro}\ \emph {et~al.}(1976)\citenamefont {de~Alfaro}, \citenamefont {Fubini},\ and\ \citenamefont {Furlan}}]{deAlfaro:1976qet}%
  \BibitemOpen
  \bibfield  {author} {\bibinfo {author} {\bibfnamefont {V.}~\bibnamefont {de~Alfaro}}, \bibinfo {author} {\bibfnamefont {S.}~\bibnamefont {Fubini}}, \ and\ \bibinfo {author} {\bibfnamefont {G.}~\bibnamefont {Furlan}},\ }\href {\doibase 10.1016/0370-2693(76)90022-8} {\bibfield  {journal} {\bibinfo  {journal} {Phys. Lett. B}\ }\textbf {\bibinfo {volume} {65}},\ \bibinfo {pages} {163} (\bibinfo {year} {1976})}\BibitemShut {NoStop}%
\bibitem [{\citenamefont {Diez}\ and\ \citenamefont {Guajardo}(2026)}]{Diez:2026bmp}%
  \BibitemOpen
  \bibfield  {author} {\bibinfo {author} {\bibfnamefont {B.}~\bibnamefont {Diez}}\ and\ \bibinfo {author} {\bibfnamefont {L.}~\bibnamefont {Guajardo}},\ }\href@noop {} {\  (\bibinfo {year} {2026})},\ \Eprint {http://arxiv.org/abs/2604.14046} {arXiv:2604.14046 [hep-th]} \BibitemShut {NoStop}%
\bibitem [{\citenamefont {Callan}\ \emph {et~al.}(1977)\citenamefont {Callan}, \citenamefont {Dashen},\ and\ \citenamefont {Gross}}]{Callan:1977qs}%
  \BibitemOpen
  \bibfield  {author} {\bibinfo {author} {\bibfnamefont {C.~G.}\ \bibnamefont {Callan}, \bibfnamefont {Jr.}}, \bibinfo {author} {\bibfnamefont {R.~F.}\ \bibnamefont {Dashen}}, \ and\ \bibinfo {author} {\bibfnamefont {D.~J.}\ \bibnamefont {Gross}},\ }\href {\doibase 10.1016/0370-2693(77)90019-3} {\bibfield  {journal} {\bibinfo  {journal} {Phys. Lett. B}\ }\textbf {\bibinfo {volume} {66}},\ \bibinfo {pages} {375} (\bibinfo {year} {1977})}\BibitemShut {NoStop}%
\bibitem [{\citenamefont {Callan}\ \emph {et~al.}(1978)\citenamefont {Callan}, \citenamefont {Dashen},\ and\ \citenamefont {Gross}}]{Callan:1977gz}%
  \BibitemOpen
  \bibfield  {author} {\bibinfo {author} {\bibfnamefont {C.~G.}\ \bibnamefont {Callan}, \bibfnamefont {Jr.}}, \bibinfo {author} {\bibfnamefont {R.~F.}\ \bibnamefont {Dashen}}, \ and\ \bibinfo {author} {\bibfnamefont {D.~J.}\ \bibnamefont {Gross}},\ }\href {\doibase 10.1103/PhysRevD.17.2717} {\bibfield  {journal} {\bibinfo  {journal} {Phys. Rev. D}\ }\textbf {\bibinfo {volume} {17}},\ \bibinfo {pages} {2717} (\bibinfo {year} {1978})}\BibitemShut {NoStop}%
\bibitem [{\citenamefont {Callan}\ \emph {et~al.}(1979)\citenamefont {Callan}, \citenamefont {Dashen},\ and\ \citenamefont {Gross}}]{Callan:1978bm}%
  \BibitemOpen
  \bibfield  {author} {\bibinfo {author} {\bibfnamefont {C.~G.}\ \bibnamefont {Callan}, \bibfnamefont {Jr.}}, \bibinfo {author} {\bibfnamefont {R.~F.}\ \bibnamefont {Dashen}}, \ and\ \bibinfo {author} {\bibfnamefont {D.~J.}\ \bibnamefont {Gross}},\ }\href {\doibase 10.1103/PhysRevD.19.1826} {\bibfield  {journal} {\bibinfo  {journal} {Phys. Rev. D}\ }\textbf {\bibinfo {volume} {19}},\ \bibinfo {pages} {1826} (\bibinfo {year} {1979})}\BibitemShut {NoStop}%
\bibitem [{\citenamefont {Negele}(1999)}]{Negele:1998ev}%
  \BibitemOpen
  \bibfield  {author} {\bibinfo {author} {\bibfnamefont {J.~W.}\ \bibnamefont {Negele}},\ }\href {\doibase 10.1016/S0920-5632(99)85010-5} {\bibfield  {journal} {\bibinfo  {journal} {Nucl. Phys. B Proc. Suppl.}\ }\textbf {\bibinfo {volume} {73}},\ \bibinfo {pages} {92} (\bibinfo {year} {1999})},\ \Eprint {http://arxiv.org/abs/hep-lat/9810053} {arXiv:hep-lat/9810053} \BibitemShut {NoStop}%
\bibitem [{\citenamefont {Steele}\ and\ \citenamefont {Negele}(2000)}]{Steele:2000xk}%
  \BibitemOpen
  \bibfield  {author} {\bibinfo {author} {\bibfnamefont {J.~V.}\ \bibnamefont {Steele}}\ and\ \bibinfo {author} {\bibfnamefont {J.~W.}\ \bibnamefont {Negele}},\ }\href {\doibase 10.1103/PhysRevLett.85.4207} {\bibfield  {journal} {\bibinfo  {journal} {Phys. Rev. Lett.}\ }\textbf {\bibinfo {volume} {85}},\ \bibinfo {pages} {4207} (\bibinfo {year} {2000})},\ \Eprint {http://arxiv.org/abs/hep-lat/0007006} {arXiv:hep-lat/0007006} \BibitemShut {NoStop}%
\bibitem [{\citenamefont {Polyakov}(1977)}]{Polyakov:1976fu}%
  \BibitemOpen
  \bibfield  {author} {\bibinfo {author} {\bibfnamefont {A.~M.}\ \bibnamefont {Polyakov}},\ }\href {\doibase 10.1016/0550-3213(77)90086-4} {\bibfield  {journal} {\bibinfo  {journal} {Nucl. Phys. B}\ }\textbf {\bibinfo {volume} {120}},\ \bibinfo {pages} {429} (\bibinfo {year} {1977})}\BibitemShut {NoStop}%
\bibitem [{\citenamefont {Actor}(1979)}]{Actor:1979in}%
  \BibitemOpen
  \bibfield  {author} {\bibinfo {author} {\bibfnamefont {A.}~\bibnamefont {Actor}},\ }\href {\doibase 10.1103/RevModPhys.51.461} {\bibfield  {journal} {\bibinfo  {journal} {Rev. Mod. Phys.}\ }\textbf {\bibinfo {volume} {51}},\ \bibinfo {pages} {461} (\bibinfo {year} {1979})}\BibitemShut {NoStop}%
\bibitem [{\citenamefont {Canfora}\ \emph {et~al.}(2013)\citenamefont {Canfora}, \citenamefont {Correa}, \citenamefont {Giacomini},\ and\ \citenamefont {Oliva}}]{Canfora:2012ap}%
  \BibitemOpen
  \bibfield  {author} {\bibinfo {author} {\bibfnamefont {F.}~\bibnamefont {Canfora}}, \bibinfo {author} {\bibfnamefont {F.}~\bibnamefont {Correa}}, \bibinfo {author} {\bibfnamefont {A.}~\bibnamefont {Giacomini}}, \ and\ \bibinfo {author} {\bibfnamefont {J.}~\bibnamefont {Oliva}},\ }\href {\doibase 10.1016/j.physletb.2013.04.029} {\bibfield  {journal} {\bibinfo  {journal} {Phys. Lett. B}\ }\textbf {\bibinfo {volume} {722}},\ \bibinfo {pages} {364} (\bibinfo {year} {2013})},\ \Eprint {http://arxiv.org/abs/1208.6042} {arXiv:1208.6042 [hep-th]} \BibitemShut {NoStop}%
\bibitem [{\citenamefont {Canfora}\ \emph {et~al.}(2017{\natexlab{a}})\citenamefont {Canfora}, \citenamefont {Oh},\ and\ \citenamefont {Salgado-Rebolledo}}]{Canfora:2017yio}%
  \BibitemOpen
  \bibfield  {author} {\bibinfo {author} {\bibfnamefont {F.}~\bibnamefont {Canfora}}, \bibinfo {author} {\bibfnamefont {S.~H.}\ \bibnamefont {Oh}}, \ and\ \bibinfo {author} {\bibfnamefont {P.}~\bibnamefont {Salgado-Rebolledo}},\ }\href {\doibase 10.1103/PhysRevD.96.084038} {\bibfield  {journal} {\bibinfo  {journal} {Phys. Rev. D}\ }\textbf {\bibinfo {volume} {96}},\ \bibinfo {pages} {084038} (\bibinfo {year} {2017}{\natexlab{a}})},\ \Eprint {http://arxiv.org/abs/1710.00133} {arXiv:1710.00133 [hep-th]} \BibitemShut {NoStop}%
\bibitem [{\citenamefont {Canfora}\ \emph {et~al.}(2019)\citenamefont {Canfora}, \citenamefont {Gomberoff}, \citenamefont {Oh}, \citenamefont {Rojas},\ and\ \citenamefont {Salgado-Rebolledo}}]{Canfora:2018ppu}%
  \BibitemOpen
  \bibfield  {author} {\bibinfo {author} {\bibfnamefont {F.}~\bibnamefont {Canfora}}, \bibinfo {author} {\bibfnamefont {A.}~\bibnamefont {Gomberoff}}, \bibinfo {author} {\bibfnamefont {S.~H.}\ \bibnamefont {Oh}}, \bibinfo {author} {\bibfnamefont {F.}~\bibnamefont {Rojas}}, \ and\ \bibinfo {author} {\bibfnamefont {P.}~\bibnamefont {Salgado-Rebolledo}},\ }\href {\doibase 10.1007/JHEP06(2019)081} {\bibfield  {journal} {\bibinfo  {journal} {JHEP}\ }\textbf {\bibinfo {volume} {06}},\ \bibinfo {pages} {081} (\bibinfo {year} {2019})},\ \Eprint {http://arxiv.org/abs/1812.11231} {arXiv:1812.11231 [hep-th]} \BibitemShut {NoStop}%
\bibitem [{\citenamefont {Canfora}\ \emph {et~al.}(2022)\citenamefont {Canfora}, \citenamefont {Gomberoff}, \citenamefont {Lagos},\ and\ \citenamefont {Vera}}]{Canfora:2022nso}%
  \BibitemOpen
  \bibfield  {author} {\bibinfo {author} {\bibfnamefont {F.}~\bibnamefont {Canfora}}, \bibinfo {author} {\bibfnamefont {A.}~\bibnamefont {Gomberoff}}, \bibinfo {author} {\bibfnamefont {M.}~\bibnamefont {Lagos}}, \ and\ \bibinfo {author} {\bibfnamefont {A.}~\bibnamefont {Vera}},\ }\href {\doibase 10.1103/PhysRevD.105.084045} {\bibfield  {journal} {\bibinfo  {journal} {Phys. Rev. D}\ }\textbf {\bibinfo {volume} {105}},\ \bibinfo {pages} {084045} (\bibinfo {year} {2022})},\ \Eprint {http://arxiv.org/abs/2203.02365} {arXiv:2203.02365 [hep-th]} \BibitemShut {NoStop}%
\bibitem [{\citenamefont {Flores-Alfonso}\ and\ \citenamefont {Larios}(2020)}]{Flores-Alfonso:2020ayc}%
  \BibitemOpen
  \bibfield  {author} {\bibinfo {author} {\bibfnamefont {D.}~\bibnamefont {Flores-Alfonso}}\ and\ \bibinfo {author} {\bibfnamefont {B.~O.}\ \bibnamefont {Larios}},\ }\href {\doibase 10.1103/PhysRevD.102.064017} {\bibfield  {journal} {\bibinfo  {journal} {Phys. Rev. D}\ }\textbf {\bibinfo {volume} {102}},\ \bibinfo {pages} {064017} (\bibinfo {year} {2020})},\ \Eprint {http://arxiv.org/abs/2005.08437} {arXiv:2005.08437 [gr-qc]} \BibitemShut {NoStop}%
\bibitem [{\citenamefont {Canfora}\ and\ \citenamefont {Salgado-Rebolledo}(2013)}]{Canfora:2013hedgehog}%
  \BibitemOpen
  \bibfield  {author} {\bibinfo {author} {\bibfnamefont {F.}~\bibnamefont {Canfora}}\ and\ \bibinfo {author} {\bibfnamefont {P.}~\bibnamefont {Salgado-Rebolledo}},\ }\href {\doibase 10.1103/PhysRevD.87.045023} {\bibfield  {journal} {\bibinfo  {journal} {Phys. Rev. D}\ }\textbf {\bibinfo {volume} {87}},\ \bibinfo {pages} {045023} (\bibinfo {year} {2013})},\ \Eprint {http://arxiv.org/abs/1302.1264} {arXiv:1302.1264 [hep-th]} \BibitemShut {NoStop}%
\bibitem [{\citenamefont {Canfora}\ and\ \citenamefont {Maeda}(2013)}]{Canfora:2013maeda}%
  \BibitemOpen
  \bibfield  {author} {\bibinfo {author} {\bibfnamefont {F.}~\bibnamefont {Canfora}}\ and\ \bibinfo {author} {\bibfnamefont {H.}~\bibnamefont {Maeda}},\ }\href {\doibase 10.1103/PhysRevD.87.084049} {\bibfield  {journal} {\bibinfo  {journal} {Phys. Rev. D}\ }\textbf {\bibinfo {volume} {87}},\ \bibinfo {pages} {084049} (\bibinfo {year} {2013})},\ \Eprint {http://arxiv.org/abs/1302.3232} {arXiv:1302.3232 [hep-th]} \BibitemShut {NoStop}%
\bibitem [{\citenamefont {Canfora}(2013)}]{Canfora:2013crystals}%
  \BibitemOpen
  \bibfield  {author} {\bibinfo {author} {\bibfnamefont {F.}~\bibnamefont {Canfora}},\ }\href {\doibase 10.1103/PhysRevD.88.065028} {\bibfield  {journal} {\bibinfo  {journal} {Phys. Rev. D}\ }\textbf {\bibinfo {volume} {88}},\ \bibinfo {pages} {065028} (\bibinfo {year} {2013})},\ \Eprint {http://arxiv.org/abs/1307.0211} {arXiv:1307.0211 [hep-th]} \BibitemShut {NoStop}%
\bibitem [{\citenamefont {Chen}\ \emph {et~al.}(2014)\citenamefont {Chen}, \citenamefont {Li},\ and\ \citenamefont {Yang}}]{Chen:2014kink}%
  \BibitemOpen
  \bibfield  {author} {\bibinfo {author} {\bibfnamefont {S.}~\bibnamefont {Chen}}, \bibinfo {author} {\bibfnamefont {Y.-X.}\ \bibnamefont {Li}}, \ and\ \bibinfo {author} {\bibfnamefont {Y.}~\bibnamefont {Yang}},\ }\href {\doibase 10.1103/PhysRevD.89.025007} {\bibfield  {journal} {\bibinfo  {journal} {Phys. Rev. D}\ }\textbf {\bibinfo {volume} {89}},\ \bibinfo {pages} {025007} (\bibinfo {year} {2014})},\ \Eprint {http://arxiv.org/abs/1312.2479} {arXiv:1312.2479 [hep-th]} \BibitemShut {NoStop}%
\bibitem [{\citenamefont {Canfora}\ \emph {et~al.}(2014{\natexlab{a}})\citenamefont {Canfora}, \citenamefont {Correa},\ and\ \citenamefont {Zanelli}}]{Canfora:2014multisoliton}%
  \BibitemOpen
  \bibfield  {author} {\bibinfo {author} {\bibfnamefont {F.}~\bibnamefont {Canfora}}, \bibinfo {author} {\bibfnamefont {F.}~\bibnamefont {Correa}}, \ and\ \bibinfo {author} {\bibfnamefont {J.}~\bibnamefont {Zanelli}},\ }\href {\doibase 10.1103/PhysRevD.90.085002} {\bibfield  {journal} {\bibinfo  {journal} {Phys. Rev. D}\ }\textbf {\bibinfo {volume} {90}},\ \bibinfo {pages} {085002} (\bibinfo {year} {2014}{\natexlab{a}})},\ \Eprint {http://arxiv.org/abs/1406.4136} {arXiv:1406.4136 [hep-th]} \BibitemShut {NoStop}%
\bibitem [{\citenamefont {Canfora}\ \emph {et~al.}(2014{\natexlab{b}})\citenamefont {Canfora}, \citenamefont {Giacomini},\ and\ \citenamefont {Pavluchenko}}]{Canfora:2014cosmo}%
  \BibitemOpen
  \bibfield  {author} {\bibinfo {author} {\bibfnamefont {F.}~\bibnamefont {Canfora}}, \bibinfo {author} {\bibfnamefont {A.}~\bibnamefont {Giacomini}}, \ and\ \bibinfo {author} {\bibfnamefont {S.~A.}\ \bibnamefont {Pavluchenko}},\ }\href {\doibase 10.1103/PhysRevD.90.043516} {\bibfield  {journal} {\bibinfo  {journal} {Phys. Rev. D}\ }\textbf {\bibinfo {volume} {90}},\ \bibinfo {pages} {043516} (\bibinfo {year} {2014}{\natexlab{b}})},\ \Eprint {http://arxiv.org/abs/1406.1541} {arXiv:1406.1541 [gr-qc]} \BibitemShut {NoStop}%
\bibitem [{\citenamefont {Canfora}\ and\ \citenamefont {Tallarita}(2014)}]{Canfora:2014monopoles}%
  \BibitemOpen
  \bibfield  {author} {\bibinfo {author} {\bibfnamefont {F.}~\bibnamefont {Canfora}}\ and\ \bibinfo {author} {\bibfnamefont {G.}~\bibnamefont {Tallarita}},\ }\href {\doibase 10.1007/JHEP09(2014)136} {\bibfield  {journal} {\bibinfo  {journal} {JHEP}\ }\textbf {\bibinfo {volume} {09}},\ \bibinfo {pages} {136} (\bibinfo {year} {2014})},\ \Eprint {http://arxiv.org/abs/1407.0609} {arXiv:1407.0609 [hep-th]} \BibitemShut {NoStop}%
\bibitem [{\citenamefont {Canfora}\ \emph {et~al.}(2015)\citenamefont {Canfora}, \citenamefont {Di~Mauro}, \citenamefont {Kurkov},\ and\ \citenamefont {Naddeo}}]{Canfora:2015sun}%
  \BibitemOpen
  \bibfield  {author} {\bibinfo {author} {\bibfnamefont {F.}~\bibnamefont {Canfora}}, \bibinfo {author} {\bibfnamefont {M.}~\bibnamefont {Di~Mauro}}, \bibinfo {author} {\bibfnamefont {M.~A.}\ \bibnamefont {Kurkov}}, \ and\ \bibinfo {author} {\bibfnamefont {A.}~\bibnamefont {Naddeo}},\ }\href {\doibase 10.1140/epjc/s10052-015-3670-5} {\bibfield  {journal} {\bibinfo  {journal} {Eur. Phys. J. C}\ }\textbf {\bibinfo {volume} {75}},\ \bibinfo {pages} {443} (\bibinfo {year} {2015})},\ \Eprint {http://arxiv.org/abs/1508.07055} {arXiv:1508.07055 [hep-th]} \BibitemShut {NoStop}%
\bibitem [{\citenamefont {Chen}\ and\ \citenamefont {Yang}(2016)}]{Chen:2016domain}%
  \BibitemOpen
  \bibfield  {author} {\bibinfo {author} {\bibfnamefont {S.}~\bibnamefont {Chen}}\ and\ \bibinfo {author} {\bibfnamefont {Y.}~\bibnamefont {Yang}},\ }\href {\doibase 10.1016/j.nuclphysb.2016.01.012} {\bibfield  {journal} {\bibinfo  {journal} {Nucl. Phys. B}\ }\textbf {\bibinfo {volume} {904}},\ \bibinfo {pages} {470} (\bibinfo {year} {2016})},\ \Eprint {http://arxiv.org/abs/1512.04080} {arXiv:1512.04080 [hep-th]} \BibitemShut {NoStop}%
\bibitem [{\citenamefont {Ayón-Beato}\ \emph {et~al.}(2016)\citenamefont {Ayón-Beato}, \citenamefont {Canfora},\ and\ \citenamefont {Zanelli}}]{AyonBeato:2016skyrmions}%
  \BibitemOpen
  \bibfield  {author} {\bibinfo {author} {\bibfnamefont {E.}~\bibnamefont {Ayón-Beato}}, \bibinfo {author} {\bibfnamefont {F.}~\bibnamefont {Canfora}}, \ and\ \bibinfo {author} {\bibfnamefont {J.}~\bibnamefont {Zanelli}},\ }\href {\doibase 10.1016/j.physletb.2015.11.043} {\bibfield  {journal} {\bibinfo  {journal} {Phys. Lett. B}\ }\textbf {\bibinfo {volume} {752}},\ \bibinfo {pages} {201} (\bibinfo {year} {2016})},\ \Eprint {http://arxiv.org/abs/1509.02659} {arXiv:1509.02659 [gr-qc]} \BibitemShut {NoStop}%
\bibitem [{\citenamefont {Canfora}\ and\ \citenamefont {Tallarita}(2015)}]{Canfora:2015bps}%
  \BibitemOpen
  \bibfield  {author} {\bibinfo {author} {\bibfnamefont {F.}~\bibnamefont {Canfora}}\ and\ \bibinfo {author} {\bibfnamefont {G.}~\bibnamefont {Tallarita}},\ }\href {\doibase 10.1103/PhysRevD.91.085033} {\bibfield  {journal} {\bibinfo  {journal} {Phys. Rev. D}\ }\textbf {\bibinfo {volume} {91}},\ \bibinfo {pages} {085033} (\bibinfo {year} {2015})},\ \Eprint {http://arxiv.org/abs/1502.02957} {arXiv:1502.02957 [hep-th]} \BibitemShut {NoStop}%
\bibitem [{\citenamefont {Canfora}\ and\ \citenamefont {Tallarita}(2016)}]{Canfora:2016orientational}%
  \BibitemOpen
  \bibfield  {author} {\bibinfo {author} {\bibfnamefont {F.}~\bibnamefont {Canfora}}\ and\ \bibinfo {author} {\bibfnamefont {G.}~\bibnamefont {Tallarita}},\ }\href {\doibase 10.1103/PhysRevD.94.025037} {\bibfield  {journal} {\bibinfo  {journal} {Phys. Rev. D}\ }\textbf {\bibinfo {volume} {94}},\ \bibinfo {pages} {025037} (\bibinfo {year} {2016})},\ \Eprint {http://arxiv.org/abs/1607.04140} {arXiv:1607.04140 [hep-th]} \BibitemShut {NoStop}%
\bibitem [{\citenamefont {Tallarita}\ and\ \citenamefont {Canfora}(2017)}]{Tallarita:2017popcorn}%
  \BibitemOpen
  \bibfield  {author} {\bibinfo {author} {\bibfnamefont {G.}~\bibnamefont {Tallarita}}\ and\ \bibinfo {author} {\bibfnamefont {F.}~\bibnamefont {Canfora}},\ }\href {\doibase 10.1016/j.nuclphysb.2017.05.015} {\bibfield  {journal} {\bibinfo  {journal} {Nucl. Phys. B}\ }\textbf {\bibinfo {volume} {921}},\ \bibinfo {pages} {394} (\bibinfo {year} {2017})},\ \Eprint {http://arxiv.org/abs/1706.01397} {arXiv:1706.01397 [hep-th]} \BibitemShut {NoStop}%
\bibitem [{\citenamefont {Canfora}\ \emph {et~al.}(2017{\natexlab{b}})\citenamefont {Canfora}, \citenamefont {Paliathanasis}, \citenamefont {Taves},\ and\ \citenamefont {Zanelli}}]{Canfora:2017einsteinskyrme}%
  \BibitemOpen
  \bibfield  {author} {\bibinfo {author} {\bibfnamefont {F.}~\bibnamefont {Canfora}}, \bibinfo {author} {\bibfnamefont {A.}~\bibnamefont {Paliathanasis}}, \bibinfo {author} {\bibfnamefont {T.}~\bibnamefont {Taves}}, \ and\ \bibinfo {author} {\bibfnamefont {J.}~\bibnamefont {Zanelli}},\ }\href {\doibase 10.1103/PhysRevD.95.065032} {\bibfield  {journal} {\bibinfo  {journal} {Phys. Rev. D}\ }\textbf {\bibinfo {volume} {95}},\ \bibinfo {pages} {065032} (\bibinfo {year} {2017}{\natexlab{b}})},\ \Eprint {http://arxiv.org/abs/1703.04860} {arXiv:1703.04860 [gr-qc]} \BibitemShut {NoStop}%
\bibitem [{\citenamefont {Canfora}\ \emph {et~al.}(2017{\natexlab{c}})\citenamefont {Canfora}, \citenamefont {Dimakis},\ and\ \citenamefont {Paliathanasis}}]{Canfora:2017sigma}%
  \BibitemOpen
  \bibfield  {author} {\bibinfo {author} {\bibfnamefont {F.}~\bibnamefont {Canfora}}, \bibinfo {author} {\bibfnamefont {N.}~\bibnamefont {Dimakis}}, \ and\ \bibinfo {author} {\bibfnamefont {A.}~\bibnamefont {Paliathanasis}},\ }\href {\doibase 10.1103/PhysRevD.96.025021} {\bibfield  {journal} {\bibinfo  {journal} {Phys. Rev. D}\ }\textbf {\bibinfo {volume} {96}},\ \bibinfo {pages} {025021} (\bibinfo {year} {2017}{\natexlab{c}})},\ \Eprint {http://arxiv.org/abs/1707.02270} {arXiv:1707.02270 [hep-th]} \BibitemShut {NoStop}%
\bibitem [{\citenamefont {Canfora}\ \emph {et~al.}(2017{\natexlab{d}})\citenamefont {Canfora}, \citenamefont {Oh},\ and\ \citenamefont {Salgado-Rebolledo}}]{Canfora:2017merons}%
  \BibitemOpen
  \bibfield  {author} {\bibinfo {author} {\bibfnamefont {F.}~\bibnamefont {Canfora}}, \bibinfo {author} {\bibfnamefont {S.~H.}\ \bibnamefont {Oh}}, \ and\ \bibinfo {author} {\bibfnamefont {P.}~\bibnamefont {Salgado-Rebolledo}},\ }\href {\doibase 10.1103/PhysRevD.96.084038} {\bibfield  {journal} {\bibinfo  {journal} {Phys. Rev. D}\ }\textbf {\bibinfo {volume} {96}},\ \bibinfo {pages} {084038} (\bibinfo {year} {2017}{\natexlab{d}})},\ \Eprint {http://arxiv.org/abs/1710.00133} {arXiv:1710.00133 [hep-th]} \BibitemShut {NoStop}%
\bibitem [{\citenamefont {Canfora}\ \emph {et~al.}(2025{\natexlab{a}})\citenamefont {Canfora}, \citenamefont {Corral},\ and\ \citenamefont {Diez}}]{Canfora:2025roy}%
  \BibitemOpen
  \bibfield  {author} {\bibinfo {author} {\bibfnamefont {F.}~\bibnamefont {Canfora}}, \bibinfo {author} {\bibfnamefont {C.}~\bibnamefont {Corral}}, \ and\ \bibinfo {author} {\bibfnamefont {B.}~\bibnamefont {Diez}},\ }\href {\doibase 10.1103/PhysRevD.111.084072} {\bibfield  {journal} {\bibinfo  {journal} {Phys. Rev. D}\ }\textbf {\bibinfo {volume} {111}},\ \bibinfo {pages} {084072} (\bibinfo {year} {2025}{\natexlab{a}})},\ \Eprint {http://arxiv.org/abs/2501.13024} {arXiv:2501.13024 [hep-th]} \BibitemShut {NoStop}%
\bibitem [{\citenamefont {Ipinza}\ and\ \citenamefont {Salgado-Rebolledo}(2021)}]{Ipinza:2020xgc}%
  \BibitemOpen
  \bibfield  {author} {\bibinfo {author} {\bibfnamefont {M.}~\bibnamefont {Ipinza}}\ and\ \bibinfo {author} {\bibfnamefont {P.}~\bibnamefont {Salgado-Rebolledo}},\ }\href {\doibase 10.1140/epjc/s10052-021-09444-7} {\bibfield  {journal} {\bibinfo  {journal} {Eur. Phys. J. C}\ }\textbf {\bibinfo {volume} {81}},\ \bibinfo {pages} {654} (\bibinfo {year} {2021})},\ \Eprint {http://arxiv.org/abs/2005.04920} {arXiv:2005.04920 [hep-th]} \BibitemShut {NoStop}%
\bibitem [{\citenamefont {Bahamonde}(2026)}]{Bahamonde:2026bvh}%
  \BibitemOpen
  \bibfield  {author} {\bibinfo {author} {\bibfnamefont {S.}~\bibnamefont {Bahamonde}},\ }\href@noop {} {\  (\bibinfo {year} {2026})},\ \Eprint {http://arxiv.org/abs/2604.15758} {arXiv:2604.15758 [gr-qc]} \BibitemShut {NoStop}%
\bibitem [{\citenamefont {Herdeiro}\ and\ \citenamefont {Radu}(2015)}]{Herdeiro:2015waa}%
  \BibitemOpen
  \bibfield  {author} {\bibinfo {author} {\bibfnamefont {C.~A.~R.}\ \bibnamefont {Herdeiro}}\ and\ \bibinfo {author} {\bibfnamefont {E.}~\bibnamefont {Radu}},\ }\href {\doibase 10.1142/S0218271815420146} {\bibfield  {journal} {\bibinfo  {journal} {Int. J. Mod. Phys. D}\ }\textbf {\bibinfo {volume} {24}},\ \bibinfo {pages} {1542014} (\bibinfo {year} {2015})},\ \Eprint {http://arxiv.org/abs/1504.08209} {arXiv:1504.08209 [gr-qc]} \BibitemShut {NoStop}%
\bibitem [{\citenamefont {Bekenstein}(1974)}]{Bekenstein:1974sf}%
  \BibitemOpen
  \bibfield  {author} {\bibinfo {author} {\bibfnamefont {J.~D.}\ \bibnamefont {Bekenstein}},\ }\href {\doibase 10.1016/0003-4916(74)90124-9} {\bibfield  {journal} {\bibinfo  {journal} {Annals Phys.}\ }\textbf {\bibinfo {volume} {82}},\ \bibinfo {pages} {535} (\bibinfo {year} {1974})}\BibitemShut {NoStop}%
\bibitem [{\citenamefont {Martinez}\ \emph {et~al.}(2003)\citenamefont {Martinez}, \citenamefont {Troncoso},\ and\ \citenamefont {Zanelli}}]{Martinez:2002ru}%
  \BibitemOpen
  \bibfield  {author} {\bibinfo {author} {\bibfnamefont {C.}~\bibnamefont {Martinez}}, \bibinfo {author} {\bibfnamefont {R.}~\bibnamefont {Troncoso}}, \ and\ \bibinfo {author} {\bibfnamefont {J.}~\bibnamefont {Zanelli}},\ }\href {\doibase 10.1103/PhysRevD.67.024008} {\bibfield  {journal} {\bibinfo  {journal} {Phys. Rev. D}\ }\textbf {\bibinfo {volume} {67}},\ \bibinfo {pages} {024008} (\bibinfo {year} {2003})},\ \Eprint {http://arxiv.org/abs/hep-th/0205319} {arXiv:hep-th/0205319} \BibitemShut {NoStop}%
\bibitem [{\citenamefont {Martinez}\ \emph {et~al.}(2006)\citenamefont {Martinez}, \citenamefont {Staforelli},\ and\ \citenamefont {Troncoso}}]{Martinez:2005di}%
  \BibitemOpen
  \bibfield  {author} {\bibinfo {author} {\bibfnamefont {C.}~\bibnamefont {Martinez}}, \bibinfo {author} {\bibfnamefont {J.~P.}\ \bibnamefont {Staforelli}}, \ and\ \bibinfo {author} {\bibfnamefont {R.}~\bibnamefont {Troncoso}},\ }\href {\doibase 10.1103/PhysRevD.74.044028} {\bibfield  {journal} {\bibinfo  {journal} {Phys. Rev. D}\ }\textbf {\bibinfo {volume} {74}},\ \bibinfo {pages} {044028} (\bibinfo {year} {2006})},\ \Eprint {http://arxiv.org/abs/hep-th/0512022} {arXiv:hep-th/0512022} \BibitemShut {NoStop}%
\bibitem [{\citenamefont {Charmousis}\ \emph {et~al.}(2009)\citenamefont {Charmousis}, \citenamefont {Kolyvaris},\ and\ \citenamefont {Papantonopoulos}}]{Charmousis:2009cm}%
  \BibitemOpen
  \bibfield  {author} {\bibinfo {author} {\bibfnamefont {C.}~\bibnamefont {Charmousis}}, \bibinfo {author} {\bibfnamefont {T.}~\bibnamefont {Kolyvaris}}, \ and\ \bibinfo {author} {\bibfnamefont {E.}~\bibnamefont {Papantonopoulos}},\ }\href {\doibase 10.1088/0264-9381/26/17/175012} {\bibfield  {journal} {\bibinfo  {journal} {Class. Quant. Grav.}\ }\textbf {\bibinfo {volume} {26}},\ \bibinfo {pages} {175012} (\bibinfo {year} {2009})},\ \Eprint {http://arxiv.org/abs/0906.5568} {arXiv:0906.5568 [gr-qc]} \BibitemShut {NoStop}%
\bibitem [{\citenamefont {Bardoux}\ \emph {et~al.}(2014)\citenamefont {Bardoux}, \citenamefont {Caldarelli},\ and\ \citenamefont {Charmousis}}]{Bardoux:2013swa}%
  \BibitemOpen
  \bibfield  {author} {\bibinfo {author} {\bibfnamefont {Y.}~\bibnamefont {Bardoux}}, \bibinfo {author} {\bibfnamefont {M.~M.}\ \bibnamefont {Caldarelli}}, \ and\ \bibinfo {author} {\bibfnamefont {C.}~\bibnamefont {Charmousis}},\ }\href {\doibase 10.1007/JHEP05(2014)039} {\bibfield  {journal} {\bibinfo  {journal} {JHEP}\ }\textbf {\bibinfo {volume} {05}},\ \bibinfo {pages} {039} (\bibinfo {year} {2014})},\ \Eprint {http://arxiv.org/abs/1311.1192} {arXiv:1311.1192 [hep-th]} \BibitemShut {NoStop}%
\bibitem [{\citenamefont {Astorino}(2015)}]{Astorino:2014mda}%
  \BibitemOpen
  \bibfield  {author} {\bibinfo {author} {\bibfnamefont {M.}~\bibnamefont {Astorino}},\ }\href {\doibase 10.1103/PhysRevD.91.064066} {\bibfield  {journal} {\bibinfo  {journal} {Phys. Rev. D}\ }\textbf {\bibinfo {volume} {91}},\ \bibinfo {pages} {064066} (\bibinfo {year} {2015})},\ \Eprint {http://arxiv.org/abs/1412.3539} {arXiv:1412.3539 [gr-qc]} \BibitemShut {NoStop}%
\bibitem [{\citenamefont {Anabalon}\ and\ \citenamefont {Maeda}(2010)}]{Anabalon:2009qt}%
  \BibitemOpen
  \bibfield  {author} {\bibinfo {author} {\bibfnamefont {A.}~\bibnamefont {Anabalon}}\ and\ \bibinfo {author} {\bibfnamefont {H.}~\bibnamefont {Maeda}},\ }\href {\doibase 10.1103/PhysRevD.81.041501} {\bibfield  {journal} {\bibinfo  {journal} {Phys. Rev. D}\ }\textbf {\bibinfo {volume} {81}},\ \bibinfo {pages} {041501} (\bibinfo {year} {2010})},\ \Eprint {http://arxiv.org/abs/0907.0219} {arXiv:0907.0219 [hep-th]} \BibitemShut {NoStop}%
\bibitem [{\citenamefont {Cisterna}\ \emph {et~al.}(2021)\citenamefont {Cisterna}, \citenamefont {Neira-Gallegos}, \citenamefont {Oliva},\ and\ \citenamefont {Rebolledo-Caceres}}]{Cisterna:2021xxq}%
  \BibitemOpen
  \bibfield  {author} {\bibinfo {author} {\bibfnamefont {A.}~\bibnamefont {Cisterna}}, \bibinfo {author} {\bibfnamefont {A.}~\bibnamefont {Neira-Gallegos}}, \bibinfo {author} {\bibfnamefont {J.}~\bibnamefont {Oliva}}, \ and\ \bibinfo {author} {\bibfnamefont {S.~C.}\ \bibnamefont {Rebolledo-Caceres}},\ }\href {\doibase 10.1103/PhysRevD.103.064050} {\bibfield  {journal} {\bibinfo  {journal} {Phys. Rev. D}\ }\textbf {\bibinfo {volume} {103}},\ \bibinfo {pages} {064050} (\bibinfo {year} {2021})},\ \Eprint {http://arxiv.org/abs/2101.03628} {arXiv:2101.03628 [gr-qc]} \BibitemShut {NoStop}%
\bibitem [{\citenamefont {Caceres}\ \emph {et~al.}(2020)\citenamefont {Caceres}, \citenamefont {Figueroa}, \citenamefont {Oliva}, \citenamefont {Oyarzo},\ and\ \citenamefont {Stuardo}}]{Caceres:2020myr}%
  \BibitemOpen
  \bibfield  {author} {\bibinfo {author} {\bibfnamefont {N.}~\bibnamefont {Caceres}}, \bibinfo {author} {\bibfnamefont {J.}~\bibnamefont {Figueroa}}, \bibinfo {author} {\bibfnamefont {J.}~\bibnamefont {Oliva}}, \bibinfo {author} {\bibfnamefont {M.}~\bibnamefont {Oyarzo}}, \ and\ \bibinfo {author} {\bibfnamefont {R.}~\bibnamefont {Stuardo}},\ }\href {\doibase 10.1007/JHEP04(2020)157} {\bibfield  {journal} {\bibinfo  {journal} {JHEP}\ }\textbf {\bibinfo {volume} {04}},\ \bibinfo {pages} {157} (\bibinfo {year} {2020})},\ \Eprint {http://arxiv.org/abs/2001.01478} {arXiv:2001.01478 [hep-th]} \BibitemShut {NoStop}%
\bibitem [{\citenamefont {Barrientos}\ \emph {et~al.}(2022)\citenamefont {Barrientos}, \citenamefont {Cisterna}, \citenamefont {Mora},\ and\ \citenamefont {Vigan{\`o}}}]{Barrientos:2022avi}%
  \BibitemOpen
  \bibfield  {author} {\bibinfo {author} {\bibfnamefont {J.}~\bibnamefont {Barrientos}}, \bibinfo {author} {\bibfnamefont {A.}~\bibnamefont {Cisterna}}, \bibinfo {author} {\bibfnamefont {N.}~\bibnamefont {Mora}}, \ and\ \bibinfo {author} {\bibfnamefont {A.}~\bibnamefont {Vigan{\`o}}},\ }\href {\doibase 10.1103/PhysRevD.106.024038} {\bibfield  {journal} {\bibinfo  {journal} {Phys. Rev. D}\ }\textbf {\bibinfo {volume} {106}},\ \bibinfo {pages} {024038} (\bibinfo {year} {2022})},\ \Eprint {http://arxiv.org/abs/2202.06706} {arXiv:2202.06706 [hep-th]} \BibitemShut {NoStop}%
\bibitem [{\citenamefont {Avil{\'e}s}\ \emph {et~al.}(2018)\citenamefont {Avil{\'e}s}, \citenamefont {Maeda},\ and\ \citenamefont {Martinez}}]{Aviles:2018vnf}%
  \BibitemOpen
  \bibfield  {author} {\bibinfo {author} {\bibfnamefont {L.}~\bibnamefont {Avil{\'e}s}}, \bibinfo {author} {\bibfnamefont {H.}~\bibnamefont {Maeda}}, \ and\ \bibinfo {author} {\bibfnamefont {C.}~\bibnamefont {Martinez}},\ }\href {\doibase 10.1088/1361-6382/aaea9f} {\bibfield  {journal} {\bibinfo  {journal} {Class. Quant. Grav.}\ }\textbf {\bibinfo {volume} {35}},\ \bibinfo {pages} {245001} (\bibinfo {year} {2018})},\ \Eprint {http://arxiv.org/abs/1808.10040} {arXiv:1808.10040 [gr-qc]} \BibitemShut {NoStop}%
\bibitem [{\citenamefont {Cisterna}\ \emph {et~al.}(2023)\citenamefont {Cisterna}, \citenamefont {M{\"u}ller}, \citenamefont {Pallikaris},\ and\ \citenamefont {Vigan{\`o}}}]{Cisterna:2023uqf}%
  \BibitemOpen
  \bibfield  {author} {\bibinfo {author} {\bibfnamefont {A.}~\bibnamefont {Cisterna}}, \bibinfo {author} {\bibfnamefont {K.}~\bibnamefont {M{\"u}ller}}, \bibinfo {author} {\bibfnamefont {K.}~\bibnamefont {Pallikaris}}, \ and\ \bibinfo {author} {\bibfnamefont {A.}~\bibnamefont {Vigan{\`o}}},\ }\href {\doibase 10.1103/PhysRevD.108.024066} {\bibfield  {journal} {\bibinfo  {journal} {Phys. Rev. D}\ }\textbf {\bibinfo {volume} {108}},\ \bibinfo {pages} {024066} (\bibinfo {year} {2023})},\ \Eprint {http://arxiv.org/abs/2306.14541} {arXiv:2306.14541 [gr-qc]} \BibitemShut {NoStop}%
\bibitem [{\citenamefont {Bravo-Gaete}\ \emph {et~al.}(2026)\citenamefont {Bravo-Gaete}, \citenamefont {Santos}, \citenamefont {Herrera-Mendoza},\ and\ \citenamefont {Higuita-Borja}}]{Bravo-Gaete:2025vyd}%
  \BibitemOpen
  \bibfield  {author} {\bibinfo {author} {\bibfnamefont {M.}~\bibnamefont {Bravo-Gaete}}, \bibinfo {author} {\bibfnamefont {F.~F.}\ \bibnamefont {Santos}}, \bibinfo {author} {\bibfnamefont {J.~A.}\ \bibnamefont {Herrera-Mendoza}}, \ and\ \bibinfo {author} {\bibfnamefont {D.~F.}\ \bibnamefont {Higuita-Borja}},\ }\href {\doibase 10.1088/1361-6382/ae5866} {\bibfield  {journal} {\bibinfo  {journal} {Class. Quant. Grav.}\ }\textbf {\bibinfo {volume} {43}},\ \bibinfo {pages} {075011} (\bibinfo {year} {2026})},\ \Eprint {http://arxiv.org/abs/2504.17081} {arXiv:2504.17081 [hep-th]} \BibitemShut {NoStop}%
\bibitem [{\citenamefont {Barrientos}\ and\ \citenamefont {Cisterna}(2023)}]{Barrientos:2023tqb}%
  \BibitemOpen
  \bibfield  {author} {\bibinfo {author} {\bibfnamefont {J.}~\bibnamefont {Barrientos}}\ and\ \bibinfo {author} {\bibfnamefont {A.}~\bibnamefont {Cisterna}},\ }\href {\doibase 10.1103/PhysRevD.108.024059} {\bibfield  {journal} {\bibinfo  {journal} {Phys. Rev. D}\ }\textbf {\bibinfo {volume} {108}},\ \bibinfo {pages} {024059} (\bibinfo {year} {2023})},\ \Eprint {http://arxiv.org/abs/2305.03765} {arXiv:2305.03765 [gr-qc]} \BibitemShut {NoStop}%
\bibitem [{\citenamefont {Arratia}\ \emph {et~al.}(2021)\citenamefont {Arratia}, \citenamefont {Corral}, \citenamefont {Figueroa},\ and\ \citenamefont {Sanhueza}}]{Arratia:2020hoy}%
  \BibitemOpen
  \bibfield  {author} {\bibinfo {author} {\bibfnamefont {E.}~\bibnamefont {Arratia}}, \bibinfo {author} {\bibfnamefont {C.}~\bibnamefont {Corral}}, \bibinfo {author} {\bibfnamefont {J.}~\bibnamefont {Figueroa}}, \ and\ \bibinfo {author} {\bibfnamefont {L.}~\bibnamefont {Sanhueza}},\ }\href {\doibase 10.1103/PhysRevD.103.064068} {\bibfield  {journal} {\bibinfo  {journal} {Phys. Rev. D}\ }\textbf {\bibinfo {volume} {103}},\ \bibinfo {pages} {064068} (\bibinfo {year} {2021})},\ \Eprint {http://arxiv.org/abs/2010.02460} {arXiv:2010.02460 [hep-th]} \BibitemShut {NoStop}%
\bibitem [{\citenamefont {Corral}\ \emph {et~al.}(2025)\citenamefont {Corral}, \citenamefont {Diez},\ and\ \citenamefont {Papantonopoulos}}]{Corral:2025npd}%
  \BibitemOpen
  \bibfield  {author} {\bibinfo {author} {\bibfnamefont {C.}~\bibnamefont {Corral}}, \bibinfo {author} {\bibfnamefont {B.}~\bibnamefont {Diez}}, \ and\ \bibinfo {author} {\bibfnamefont {E.}~\bibnamefont {Papantonopoulos}},\ }\href@noop {} {\  (\bibinfo {year} {2025})},\ \Eprint {http://arxiv.org/abs/2510.11854} {arXiv:2510.11854 [hep-th]} \BibitemShut {NoStop}%
\bibitem [{\citenamefont {de~Haro}\ \emph {et~al.}(2007)\citenamefont {de~Haro}, \citenamefont {Papadimitriou},\ and\ \citenamefont {Petkou}}]{deHaro:2006ymc}%
  \BibitemOpen
  \bibfield  {author} {\bibinfo {author} {\bibfnamefont {S.}~\bibnamefont {de~Haro}}, \bibinfo {author} {\bibfnamefont {I.}~\bibnamefont {Papadimitriou}}, \ and\ \bibinfo {author} {\bibfnamefont {A.~C.}\ \bibnamefont {Petkou}},\ }\href {\doibase 10.1103/PhysRevLett.98.231601} {\bibfield  {journal} {\bibinfo  {journal} {Phys. Rev. Lett.}\ }\textbf {\bibinfo {volume} {98}},\ \bibinfo {pages} {231601} (\bibinfo {year} {2007})},\ \Eprint {http://arxiv.org/abs/hep-th/0611315} {arXiv:hep-th/0611315} \BibitemShut {NoStop}%
\bibitem [{\citenamefont {Bocharova}\ \emph {et~al.}(1970)\citenamefont {Bocharova}, \citenamefont {Bronnikov},\ and\ \citenamefont {Melnikov}}]{Bocharova:1970}%
  \BibitemOpen
  \bibfield  {author} {\bibinfo {author} {\bibfnamefont {N.~M.}\ \bibnamefont {Bocharova}}, \bibinfo {author} {\bibfnamefont {K.~A.}\ \bibnamefont {Bronnikov}}, \ and\ \bibinfo {author} {\bibfnamefont {V.~N.}\ \bibnamefont {Melnikov}},\ }\href@noop {} {\bibfield  {journal} {\bibinfo  {journal} {Vestn. Mosk. Univ. Fiz. Astron.}\ }\textbf {\bibinfo {volume} {6}},\ \bibinfo {pages} {706} (\bibinfo {year} {1970})}\BibitemShut {NoStop}%
\bibitem [{\citenamefont {Sudarsky}\ and\ \citenamefont {Zannias}(1998)}]{Sudarsky:1997te}%
  \BibitemOpen
  \bibfield  {author} {\bibinfo {author} {\bibfnamefont {D.}~\bibnamefont {Sudarsky}}\ and\ \bibinfo {author} {\bibfnamefont {T.}~\bibnamefont {Zannias}},\ }\href {\doibase 10.1103/PhysRevD.58.087502} {\bibfield  {journal} {\bibinfo  {journal} {Phys. Rev. D}\ }\textbf {\bibinfo {volume} {58}},\ \bibinfo {pages} {087502} (\bibinfo {year} {1998})},\ \Eprint {http://arxiv.org/abs/gr-qc/9712083} {arXiv:gr-qc/9712083} \BibitemShut {NoStop}%
\bibitem [{\citenamefont {Astorino}(2014)}]{Astorino:2013xxa}%
  \BibitemOpen
  \bibfield  {author} {\bibinfo {author} {\bibfnamefont {M.}~\bibnamefont {Astorino}},\ }\href {\doibase 10.1103/PhysRevD.89.044022} {\bibfield  {journal} {\bibinfo  {journal} {Phys. Rev. D}\ }\textbf {\bibinfo {volume} {89}},\ \bibinfo {pages} {044022} (\bibinfo {year} {2014})},\ \Eprint {http://arxiv.org/abs/1312.1723} {arXiv:1312.1723 [gr-qc]} \BibitemShut {NoStop}%
\bibitem [{\citenamefont {Anastasiou}\ \emph {et~al.}(2023)\citenamefont {Anastasiou}, \citenamefont {Araya}, \citenamefont {Busnego-Barrientos}, \citenamefont {Corral},\ and\ \citenamefont {Merino}}]{Anastasiou:2022wjq}%
  \BibitemOpen
  \bibfield  {author} {\bibinfo {author} {\bibfnamefont {G.}~\bibnamefont {Anastasiou}}, \bibinfo {author} {\bibfnamefont {I.~J.}\ \bibnamefont {Araya}}, \bibinfo {author} {\bibfnamefont {M.}~\bibnamefont {Busnego-Barrientos}}, \bibinfo {author} {\bibfnamefont {C.}~\bibnamefont {Corral}}, \ and\ \bibinfo {author} {\bibfnamefont {N.}~\bibnamefont {Merino}},\ }\href {\doibase 10.1103/PhysRevD.107.104049} {\bibfield  {journal} {\bibinfo  {journal} {Phys. Rev. D}\ }\textbf {\bibinfo {volume} {107}},\ \bibinfo {pages} {104049} (\bibinfo {year} {2023})},\ \Eprint {http://arxiv.org/abs/2212.04364} {arXiv:2212.04364 [hep-th]} \BibitemShut {NoStop}%
\bibitem [{\citenamefont {Xanthopoulos}\ and\ \citenamefont {Dialynas}(1992)}]{Xanthopoulos:1992fm}%
  \BibitemOpen
  \bibfield  {author} {\bibinfo {author} {\bibfnamefont {B.~C.}\ \bibnamefont {Xanthopoulos}}\ and\ \bibinfo {author} {\bibfnamefont {T.~E.}\ \bibnamefont {Dialynas}},\ }\href {\doibase 10.1063/1.529723} {\bibfield  {journal} {\bibinfo  {journal} {J. Math. Phys.}\ }\textbf {\bibinfo {volume} {33}},\ \bibinfo {pages} {1463} (\bibinfo {year} {1992})}\BibitemShut {NoStop}%
\bibitem [{\citenamefont {Klimcik}(1993)}]{Klimcik:1993cia}%
  \BibitemOpen
  \bibfield  {author} {\bibinfo {author} {\bibfnamefont {C.}~\bibnamefont {Klimcik}},\ }\href {\doibase 10.1063/1.530146} {\bibfield  {journal} {\bibinfo  {journal} {J. Math. Phys.}\ }\textbf {\bibinfo {volume} {34}},\ \bibinfo {pages} {1914} (\bibinfo {year} {1993})}\BibitemShut {NoStop}%
\bibitem [{\citenamefont {Oliva}\ and\ \citenamefont {Ray}(2012)}]{Oliva:2011np}%
  \BibitemOpen
  \bibfield  {author} {\bibinfo {author} {\bibfnamefont {J.}~\bibnamefont {Oliva}}\ and\ \bibinfo {author} {\bibfnamefont {S.}~\bibnamefont {Ray}},\ }\href {\doibase 10.1088/0264-9381/29/20/205008} {\bibfield  {journal} {\bibinfo  {journal} {Class. Quant. Grav.}\ }\textbf {\bibinfo {volume} {29}},\ \bibinfo {pages} {205008} (\bibinfo {year} {2012})},\ \Eprint {http://arxiv.org/abs/1112.4112} {arXiv:1112.4112 [gr-qc]} \BibitemShut {NoStop}%
\bibitem [{\citenamefont {Giribet}\ \emph {et~al.}(2014)\citenamefont {Giribet}, \citenamefont {Leoni}, \citenamefont {Oliva},\ and\ \citenamefont {Ray}}]{Giribet:2014bva}%
  \BibitemOpen
  \bibfield  {author} {\bibinfo {author} {\bibfnamefont {G.}~\bibnamefont {Giribet}}, \bibinfo {author} {\bibfnamefont {M.}~\bibnamefont {Leoni}}, \bibinfo {author} {\bibfnamefont {J.}~\bibnamefont {Oliva}}, \ and\ \bibinfo {author} {\bibfnamefont {S.}~\bibnamefont {Ray}},\ }\href {\doibase 10.1103/PhysRevD.89.085040} {\bibfield  {journal} {\bibinfo  {journal} {Phys. Rev. D}\ }\textbf {\bibinfo {volume} {89}},\ \bibinfo {pages} {085040} (\bibinfo {year} {2014})},\ \Eprint {http://arxiv.org/abs/1401.4987} {arXiv:1401.4987 [hep-th]} \BibitemShut {NoStop}%
\bibitem [{\citenamefont {Chernicoff}\ \emph {et~al.}(2016)\citenamefont {Chernicoff}, \citenamefont {Galante}, \citenamefont {Giribet}, \citenamefont {Goya}, \citenamefont {Leoni}, \citenamefont {Oliva},\ and\ \citenamefont {Perez-Nadal}}]{Chernicoff:2016jsu}%
  \BibitemOpen
  \bibfield  {author} {\bibinfo {author} {\bibfnamefont {M.}~\bibnamefont {Chernicoff}}, \bibinfo {author} {\bibfnamefont {M.}~\bibnamefont {Galante}}, \bibinfo {author} {\bibfnamefont {G.}~\bibnamefont {Giribet}}, \bibinfo {author} {\bibfnamefont {A.}~\bibnamefont {Goya}}, \bibinfo {author} {\bibfnamefont {M.}~\bibnamefont {Leoni}}, \bibinfo {author} {\bibfnamefont {J.}~\bibnamefont {Oliva}}, \ and\ \bibinfo {author} {\bibfnamefont {G.}~\bibnamefont {Perez-Nadal}},\ }\href {\doibase 10.1007/JHEP06(2016)159} {\bibfield  {journal} {\bibinfo  {journal} {JHEP}\ }\textbf {\bibinfo {volume} {06}},\ \bibinfo {pages} {159} (\bibinfo {year} {2016})},\ \Eprint {http://arxiv.org/abs/1604.08203} {arXiv:1604.08203 [hep-th]} \BibitemShut {NoStop}%
\bibitem [{\citenamefont {Babichev}\ \emph {et~al.}(2023{\natexlab{a}})\citenamefont {Babichev}, \citenamefont {Charmousis}, \citenamefont {Hassaine},\ and\ \citenamefont {Lecoeur}}]{Babichev:2023rhn}%
  \BibitemOpen
  \bibfield  {author} {\bibinfo {author} {\bibfnamefont {E.}~\bibnamefont {Babichev}}, \bibinfo {author} {\bibfnamefont {C.}~\bibnamefont {Charmousis}}, \bibinfo {author} {\bibfnamefont {M.}~\bibnamefont {Hassaine}}, \ and\ \bibinfo {author} {\bibfnamefont {N.}~\bibnamefont {Lecoeur}},\ }\href {\doibase 10.1103/PhysRevD.107.084050} {\bibfield  {journal} {\bibinfo  {journal} {Phys. Rev. D}\ }\textbf {\bibinfo {volume} {107}},\ \bibinfo {pages} {084050} (\bibinfo {year} {2023}{\natexlab{a}})},\ \Eprint {http://arxiv.org/abs/2302.02920} {arXiv:2302.02920 [gr-qc]} \BibitemShut {NoStop}%
\bibitem [{\citenamefont {Ayon-Beato}\ \emph {et~al.}(2006)\citenamefont {Ayon-Beato}, \citenamefont {Martinez},\ and\ \citenamefont {Zanelli}}]{Ayon-Beato:2004nzi}%
  \BibitemOpen
  \bibfield  {author} {\bibinfo {author} {\bibfnamefont {E.}~\bibnamefont {Ayon-Beato}}, \bibinfo {author} {\bibfnamefont {C.}~\bibnamefont {Martinez}}, \ and\ \bibinfo {author} {\bibfnamefont {J.}~\bibnamefont {Zanelli}},\ }\href {\doibase 10.1007/s10714-005-0213-x} {\bibfield  {journal} {\bibinfo  {journal} {Gen. Rel. Grav.}\ }\textbf {\bibinfo {volume} {38}},\ \bibinfo {pages} {145} (\bibinfo {year} {2006})},\ \Eprint {http://arxiv.org/abs/hep-th/0403228} {arXiv:hep-th/0403228} \BibitemShut {NoStop}%
\bibitem [{\citenamefont {Ayon-Beato}\ \emph {et~al.}(2005)\citenamefont {Ayon-Beato}, \citenamefont {Martinez}, \citenamefont {Troncoso},\ and\ \citenamefont {Zanelli}}]{Ayon-Beato:2005yoq}%
  \BibitemOpen
  \bibfield  {author} {\bibinfo {author} {\bibfnamefont {E.}~\bibnamefont {Ayon-Beato}}, \bibinfo {author} {\bibfnamefont {C.}~\bibnamefont {Martinez}}, \bibinfo {author} {\bibfnamefont {R.}~\bibnamefont {Troncoso}}, \ and\ \bibinfo {author} {\bibfnamefont {J.}~\bibnamefont {Zanelli}},\ }\href {\doibase 10.1103/PhysRevD.71.104037} {\bibfield  {journal} {\bibinfo  {journal} {Phys. Rev. D}\ }\textbf {\bibinfo {volume} {71}},\ \bibinfo {pages} {104037} (\bibinfo {year} {2005})},\ \Eprint {http://arxiv.org/abs/hep-th/0505086} {arXiv:hep-th/0505086} \BibitemShut {NoStop}%
\bibitem [{\citenamefont {Anabalon}\ and\ \citenamefont {Cisterna}(2012)}]{Anabalon:2012tu}%
  \BibitemOpen
  \bibfield  {author} {\bibinfo {author} {\bibfnamefont {A.}~\bibnamefont {Anabalon}}\ and\ \bibinfo {author} {\bibfnamefont {A.}~\bibnamefont {Cisterna}},\ }\href {\doibase 10.1103/PhysRevD.85.084035} {\bibfield  {journal} {\bibinfo  {journal} {Phys. Rev. D}\ }\textbf {\bibinfo {volume} {85}},\ \bibinfo {pages} {084035} (\bibinfo {year} {2012})},\ \Eprint {http://arxiv.org/abs/1201.2008} {arXiv:1201.2008 [hep-th]} \BibitemShut {NoStop}%
\bibitem [{\citenamefont {Zhao}\ \emph {et~al.}(2014)\citenamefont {Zhao}, \citenamefont {Xu},\ and\ \citenamefont {Zhu}}]{Zhao:2013isa}%
  \BibitemOpen
  \bibfield  {author} {\bibinfo {author} {\bibfnamefont {L.}~\bibnamefont {Zhao}}, \bibinfo {author} {\bibfnamefont {W.}~\bibnamefont {Xu}}, \ and\ \bibinfo {author} {\bibfnamefont {B.}~\bibnamefont {Zhu}},\ }\href {\doibase 10.1088/0253-6102/61/4/12} {\bibfield  {journal} {\bibinfo  {journal} {Commun. Theor. Phys.}\ }\textbf {\bibinfo {volume} {61}},\ \bibinfo {pages} {475} (\bibinfo {year} {2014})},\ \Eprint {http://arxiv.org/abs/1305.6001} {arXiv:1305.6001 [gr-qc]} \BibitemShut {NoStop}%
\bibitem [{\citenamefont {Xu}\ \emph {et~al.}(2014)\citenamefont {Xu}, \citenamefont {Zhao},\ and\ \citenamefont {Zou}}]{Xu:2014uha}%
  \BibitemOpen
  \bibfield  {author} {\bibinfo {author} {\bibfnamefont {W.}~\bibnamefont {Xu}}, \bibinfo {author} {\bibfnamefont {L.}~\bibnamefont {Zhao}}, \ and\ \bibinfo {author} {\bibfnamefont {D.-C.}\ \bibnamefont {Zou}},\ }\href@noop {} {\  (\bibinfo {year} {2014})},\ \Eprint {http://arxiv.org/abs/1406.7153} {arXiv:1406.7153 [gr-qc]} \BibitemShut {NoStop}%
\bibitem [{\citenamefont {Ay{\'o}n-Beato}\ \emph {et~al.}(2015)\citenamefont {Ay{\'o}n-Beato}, \citenamefont {Hassa{\"\i}ne},\ and\ \citenamefont {M{\'e}ndez-Zavaleta}}]{Ayon-Beato:2015ada}%
  \BibitemOpen
  \bibfield  {author} {\bibinfo {author} {\bibfnamefont {E.}~\bibnamefont {Ay{\'o}n-Beato}}, \bibinfo {author} {\bibfnamefont {M.}~\bibnamefont {Hassa{\"\i}ne}}, \ and\ \bibinfo {author} {\bibfnamefont {J.~A.}\ \bibnamefont {M{\'e}ndez-Zavaleta}},\ }\href {\doibase 10.1103/PhysRevD.92.024048} {\bibfield  {journal} {\bibinfo  {journal} {Phys. Rev. D}\ }\textbf {\bibinfo {volume} {92}},\ \bibinfo {pages} {024048} (\bibinfo {year} {2015})},\ \bibinfo {note} {[Addendum: Phys.Rev.D 96, 049905 (2017)]},\ \Eprint {http://arxiv.org/abs/1506.02277} {arXiv:1506.02277 [hep-th]} \BibitemShut {NoStop}%
\bibitem [{\citenamefont {Fan}\ and\ \citenamefont {Lu}(2015)}]{Fan:2015tua}%
  \BibitemOpen
  \bibfield  {author} {\bibinfo {author} {\bibfnamefont {Z.-Y.}\ \bibnamefont {Fan}}\ and\ \bibinfo {author} {\bibfnamefont {H.}~\bibnamefont {Lu}},\ }\href {\doibase 10.1103/PhysRevD.92.064008} {\bibfield  {journal} {\bibinfo  {journal} {Phys. Rev. D}\ }\textbf {\bibinfo {volume} {92}},\ \bibinfo {pages} {064008} (\bibinfo {year} {2015})},\ \Eprint {http://arxiv.org/abs/1505.03557} {arXiv:1505.03557 [hep-th]} \BibitemShut {NoStop}%
\bibitem [{\citenamefont {Erices}\ and\ \citenamefont {Martinez}(2018)}]{Erices:2017izj}%
  \BibitemOpen
  \bibfield  {author} {\bibinfo {author} {\bibfnamefont {C.}~\bibnamefont {Erices}}\ and\ \bibinfo {author} {\bibfnamefont {C.}~\bibnamefont {Martinez}},\ }\href {\doibase 10.1103/PhysRevD.97.024034} {\bibfield  {journal} {\bibinfo  {journal} {Phys. Rev. D}\ }\textbf {\bibinfo {volume} {97}},\ \bibinfo {pages} {024034} (\bibinfo {year} {2018})},\ \Eprint {http://arxiv.org/abs/1707.03483} {arXiv:1707.03483 [hep-th]} \BibitemShut {NoStop}%
\bibitem [{\citenamefont {Ayon-Beato}\ and\ \citenamefont {Hassaine}(2005)}]{Ayon-Beato:2005pnc}%
  \BibitemOpen
  \bibfield  {author} {\bibinfo {author} {\bibfnamefont {E.}~\bibnamefont {Ayon-Beato}}\ and\ \bibinfo {author} {\bibfnamefont {M.}~\bibnamefont {Hassaine}},\ }\href {\doibase 10.1103/PhysRevD.71.084004} {\bibfield  {journal} {\bibinfo  {journal} {Phys. Rev. D}\ }\textbf {\bibinfo {volume} {71}},\ \bibinfo {pages} {084004} (\bibinfo {year} {2005})},\ \Eprint {http://arxiv.org/abs/hep-th/0501040} {arXiv:hep-th/0501040} \BibitemShut {NoStop}%
\bibitem [{\citenamefont {Ay{\'o}n-Beato}\ \emph {et~al.}(2018)\citenamefont {Ay{\'o}n-Beato}, \citenamefont {Ram{\'\i}rez-Baca},\ and\ \citenamefont {Terrero-Escalante}}]{Ayon-Beato:2015mxf}%
  \BibitemOpen
  \bibfield  {author} {\bibinfo {author} {\bibfnamefont {E.}~\bibnamefont {Ay{\'o}n-Beato}}, \bibinfo {author} {\bibfnamefont {P.~I.}\ \bibnamefont {Ram{\'\i}rez-Baca}}, \ and\ \bibinfo {author} {\bibfnamefont {C.~A.}\ \bibnamefont {Terrero-Escalante}},\ }\href {\doibase 10.1103/PhysRevD.97.043505} {\bibfield  {journal} {\bibinfo  {journal} {Phys. Rev. D}\ }\textbf {\bibinfo {volume} {97}},\ \bibinfo {pages} {043505} (\bibinfo {year} {2018})},\ \Eprint {http://arxiv.org/abs/1512.09375} {arXiv:1512.09375 [gr-qc]} \BibitemShut {NoStop}%
\bibitem [{\citenamefont {Erices}\ \emph {et~al.}(2025)\citenamefont {Erices}, \citenamefont {Guajardo},\ and\ \citenamefont {Lara}}]{Erices:2024iah}%
  \BibitemOpen
  \bibfield  {author} {\bibinfo {author} {\bibfnamefont {C.}~\bibnamefont {Erices}}, \bibinfo {author} {\bibfnamefont {L.}~\bibnamefont {Guajardo}}, \ and\ \bibinfo {author} {\bibfnamefont {K.}~\bibnamefont {Lara}},\ }\href {\doibase 10.1088/1475-7516/2025/03/051} {\bibfield  {journal} {\bibinfo  {journal} {JCAP}\ }\textbf {\bibinfo {volume} {03}},\ \bibinfo {pages} {051} (\bibinfo {year} {2025})},\ \Eprint {http://arxiv.org/abs/2410.13719} {arXiv:2410.13719 [gr-qc]} \BibitemShut {NoStop}%
\bibitem [{\citenamefont {Fernandes}(2021)}]{Fernandes:2021dsb}%
  \BibitemOpen
  \bibfield  {author} {\bibinfo {author} {\bibfnamefont {P.~G.~S.}\ \bibnamefont {Fernandes}},\ }\href {\doibase 10.1103/PhysRevD.103.104065} {\bibfield  {journal} {\bibinfo  {journal} {Phys. Rev. D}\ }\textbf {\bibinfo {volume} {103}},\ \bibinfo {pages} {104065} (\bibinfo {year} {2021})},\ \Eprint {http://arxiv.org/abs/2105.04687} {arXiv:2105.04687 [gr-qc]} \BibitemShut {NoStop}%
\bibitem [{\citenamefont {Ay{\'o}n-Beato}\ and\ \citenamefont {Hassaine}(2024)}]{Ayon-Beato:2023bzp}%
  \BibitemOpen
  \bibfield  {author} {\bibinfo {author} {\bibfnamefont {E.}~\bibnamefont {Ay{\'o}n-Beato}}\ and\ \bibinfo {author} {\bibfnamefont {M.}~\bibnamefont {Hassaine}},\ }\href {\doibase 10.1016/j.aop.2023.169567} {\bibfield  {journal} {\bibinfo  {journal} {Annals Phys.}\ }\textbf {\bibinfo {volume} {460}},\ \bibinfo {pages} {169567} (\bibinfo {year} {2024})},\ \Eprint {http://arxiv.org/abs/2305.09806} {arXiv:2305.09806 [hep-th]} \BibitemShut {NoStop}%
\bibitem [{\citenamefont {Ay{\'o}n-Beato}\ \emph {et~al.}(2024)\citenamefont {Ay{\'o}n-Beato}, \citenamefont {Flores-Alfonso},\ and\ \citenamefont {Hassaine}}]{Ayon-Beato:2024vph}%
  \BibitemOpen
  \bibfield  {author} {\bibinfo {author} {\bibfnamefont {E.}~\bibnamefont {Ay{\'o}n-Beato}}, \bibinfo {author} {\bibfnamefont {D.}~\bibnamefont {Flores-Alfonso}}, \ and\ \bibinfo {author} {\bibfnamefont {M.}~\bibnamefont {Hassaine}},\ }\href {\doibase 10.1103/PhysRevD.110.064027} {\bibfield  {journal} {\bibinfo  {journal} {Phys. Rev. D}\ }\textbf {\bibinfo {volume} {110}},\ \bibinfo {pages} {064027} (\bibinfo {year} {2024})},\ \Eprint {http://arxiv.org/abs/2404.08753} {arXiv:2404.08753 [hep-th]} \BibitemShut {NoStop}%
\bibitem [{\citenamefont {Bertini}\ \emph {et~al.}(2006)\citenamefont {Bertini}, \citenamefont {Cacciatori},\ and\ \citenamefont {Cerchiai}}]{Bertini:2005rc}%
  \BibitemOpen
  \bibfield  {author} {\bibinfo {author} {\bibfnamefont {S.}~\bibnamefont {Bertini}}, \bibinfo {author} {\bibfnamefont {S.~L.}\ \bibnamefont {Cacciatori}}, \ and\ \bibinfo {author} {\bibfnamefont {B.~L.}\ \bibnamefont {Cerchiai}},\ }\href {\doibase 10.1063/1.2190898} {\bibfield  {journal} {\bibinfo  {journal} {J. Math. Phys.}\ }\textbf {\bibinfo {volume} {47}},\ \bibinfo {pages} {043510} (\bibinfo {year} {2006})},\ \Eprint {http://arxiv.org/abs/math-ph/0510075} {arXiv:math-ph/0510075} \BibitemShut {NoStop}%
\bibitem [{\citenamefont {Cacciatori}\ \emph {et~al.}(2017)\citenamefont {Cacciatori}, \citenamefont {Dalla~Piazza},\ and\ \citenamefont {Scotti}}]{Cacciatori:2012qi}%
  \BibitemOpen
  \bibfield  {author} {\bibinfo {author} {\bibfnamefont {S.~L.}\ \bibnamefont {Cacciatori}}, \bibinfo {author} {\bibfnamefont {F.}~\bibnamefont {Dalla~Piazza}}, \ and\ \bibinfo {author} {\bibfnamefont {A.}~\bibnamefont {Scotti}},\ }\href {\doibase 10.1090/tran/6795} {\bibfield  {journal} {\bibinfo  {journal} {Trans. Am. Math. Soc.}\ }\textbf {\bibinfo {volume} {369}},\ \bibinfo {pages} {4709} (\bibinfo {year} {2017})},\ \Eprint {http://arxiv.org/abs/1207.1262} {arXiv:1207.1262 [math.GR]} \BibitemShut {NoStop}%
\bibitem [{\citenamefont {Tilma}\ and\ \citenamefont {Sudarshan}(2004)}]{Tilma:2004kp}%
  \BibitemOpen
  \bibfield  {author} {\bibinfo {author} {\bibfnamefont {T.~E.}\ \bibnamefont {Tilma}}\ and\ \bibinfo {author} {\bibfnamefont {G.}~\bibnamefont {Sudarshan}},\ }\href {\doibase 10.1016/j.geomphys.2004.03.003} {\bibfield  {journal} {\bibinfo  {journal} {J. Geom. Phys.}\ }\textbf {\bibinfo {volume} {52}},\ \bibinfo {pages} {263} (\bibinfo {year} {2004})},\ \Eprint {http://arxiv.org/abs/math-ph/0210057} {arXiv:math-ph/0210057} \BibitemShut {NoStop}%
\bibitem [{\citenamefont {Corral}\ \emph {et~al.}(2024)\citenamefont {Corral}, \citenamefont {Flores-Alfonso}, \citenamefont {Giribet},\ and\ \citenamefont {Oliva}}]{Corral:2024xfv}%
  \BibitemOpen
  \bibfield  {author} {\bibinfo {author} {\bibfnamefont {C.}~\bibnamefont {Corral}}, \bibinfo {author} {\bibfnamefont {D.}~\bibnamefont {Flores-Alfonso}}, \bibinfo {author} {\bibfnamefont {G.}~\bibnamefont {Giribet}}, \ and\ \bibinfo {author} {\bibfnamefont {J.}~\bibnamefont {Oliva}},\ }\href {\doibase 10.1140/epjc/s10052-024-13326-z} {\bibfield  {journal} {\bibinfo  {journal} {Eur. Phys. J. C}\ }\textbf {\bibinfo {volume} {84}},\ \bibinfo {pages} {959} (\bibinfo {year} {2024})},\ \Eprint {http://arxiv.org/abs/2404.15569} {arXiv:2404.15569 [hep-th]} \BibitemShut {NoStop}%
\bibitem [{\citenamefont {Bergshoeff}\ \emph {et~al.}(2009)\citenamefont {Bergshoeff}, \citenamefont {Hohm},\ and\ \citenamefont {Townsend}}]{Bergshoeff:2009hq}%
  \BibitemOpen
  \bibfield  {author} {\bibinfo {author} {\bibfnamefont {E.~A.}\ \bibnamefont {Bergshoeff}}, \bibinfo {author} {\bibfnamefont {O.}~\bibnamefont {Hohm}}, \ and\ \bibinfo {author} {\bibfnamefont {P.~K.}\ \bibnamefont {Townsend}},\ }\href {\doibase 10.1103/PhysRevLett.102.201301} {\bibfield  {journal} {\bibinfo  {journal} {Phys. Rev. Lett.}\ }\textbf {\bibinfo {volume} {102}},\ \bibinfo {pages} {201301} (\bibinfo {year} {2009})},\ \Eprint {http://arxiv.org/abs/0901.1766} {arXiv:0901.1766 [hep-th]} \BibitemShut {NoStop}%
\bibitem [{\citenamefont {Canfora}\ and\ \citenamefont {Corral}(2023)}]{Canfora:2023bug}%
  \BibitemOpen
  \bibfield  {author} {\bibinfo {author} {\bibfnamefont {F.}~\bibnamefont {Canfora}}\ and\ \bibinfo {author} {\bibfnamefont {C.}~\bibnamefont {Corral}},\ }\href {\doibase 10.1007/JHEP11(2023)146} {\bibfield  {journal} {\bibinfo  {journal} {JHEP}\ }\textbf {\bibinfo {volume} {11}},\ \bibinfo {pages} {146} (\bibinfo {year} {2023})},\ \Eprint {http://arxiv.org/abs/2309.15693} {arXiv:2309.15693 [hep-th]} \BibitemShut {NoStop}%
\bibitem [{\citenamefont {Ay{\'o}n-Beato}\ \emph {et~al.}(2020)\citenamefont {Ay{\'o}n-Beato}, \citenamefont {Canfora}, \citenamefont {Lagos}, \citenamefont {Oliva},\ and\ \citenamefont {Vera}}]{Ayon-Beato:2019tvu}%
  \BibitemOpen
  \bibfield  {author} {\bibinfo {author} {\bibfnamefont {E.}~\bibnamefont {Ay{\'o}n-Beato}}, \bibinfo {author} {\bibfnamefont {F.}~\bibnamefont {Canfora}}, \bibinfo {author} {\bibfnamefont {M.}~\bibnamefont {Lagos}}, \bibinfo {author} {\bibfnamefont {J.}~\bibnamefont {Oliva}}, \ and\ \bibinfo {author} {\bibfnamefont {A.}~\bibnamefont {Vera}},\ }\href {\doibase 10.1140/epjc/s10052-020-7926-6} {\bibfield  {journal} {\bibinfo  {journal} {Eur. Phys. J. C}\ }\textbf {\bibinfo {volume} {80}},\ \bibinfo {pages} {384} (\bibinfo {year} {2020})},\ \Eprint {http://arxiv.org/abs/1909.00540} {arXiv:1909.00540 [hep-th]} \BibitemShut {NoStop}%
\bibitem [{\citenamefont {Barriga}\ \emph {et~al.}(2025)\citenamefont {Barriga}, \citenamefont {Henr{\'\i}quez-B{\'a}ez}, \citenamefont {Sanhueza},\ and\ \citenamefont {Vera}}]{Barriga:2025mky}%
  \BibitemOpen
  \bibfield  {author} {\bibinfo {author} {\bibfnamefont {G.}~\bibnamefont {Barriga}}, \bibinfo {author} {\bibfnamefont {C.}~\bibnamefont {Henr{\'\i}quez-B{\'a}ez}}, \bibinfo {author} {\bibfnamefont {L.}~\bibnamefont {Sanhueza}}, \ and\ \bibinfo {author} {\bibfnamefont {A.}~\bibnamefont {Vera}},\ }\href@noop {} {\  (\bibinfo {year} {2025})},\ \Eprint {http://arxiv.org/abs/2511.11382} {arXiv:2511.11382 [hep-th]} \BibitemShut {NoStop}%
\bibitem [{\citenamefont {Vera}(2025)}]{Vera:2025qqz}%
  \BibitemOpen
  \bibfield  {author} {\bibinfo {author} {\bibfnamefont {A.}~\bibnamefont {Vera}},\ }\href {\doibase 10.1007/JHEP10(2025)119} {\bibfield  {journal} {\bibinfo  {journal} {JHEP}\ }\textbf {\bibinfo {volume} {10}},\ \bibinfo {pages} {119} (\bibinfo {year} {2025})},\ \Eprint {http://arxiv.org/abs/2503.03872} {arXiv:2503.03872 [hep-th]} \BibitemShut {NoStop}%
\bibitem [{\citenamefont {Henr{\'\i}quez-Baez}\ \emph {et~al.}(2025)\citenamefont {Henr{\'\i}quez-Baez}, \citenamefont {Lagos}, \citenamefont {Rodr{\'\i}guez},\ and\ \citenamefont {Vera}}]{Henriquez-Baez:2024rjb}%
  \BibitemOpen
  \bibfield  {author} {\bibinfo {author} {\bibfnamefont {C.}~\bibnamefont {Henr{\'\i}quez-Baez}}, \bibinfo {author} {\bibfnamefont {M.}~\bibnamefont {Lagos}}, \bibinfo {author} {\bibfnamefont {E.}~\bibnamefont {Rodr{\'\i}guez}}, \ and\ \bibinfo {author} {\bibfnamefont {A.}~\bibnamefont {Vera}},\ }\href {\doibase 10.1103/PhysRevD.111.104016} {\bibfield  {journal} {\bibinfo  {journal} {Phys. Rev. D}\ }\textbf {\bibinfo {volume} {111}},\ \bibinfo {pages} {104016} (\bibinfo {year} {2025})},\ \Eprint {http://arxiv.org/abs/2412.12343} {arXiv:2412.12343 [hep-th]} \BibitemShut {NoStop}%
\bibitem [{\citenamefont {Concha}\ \emph {et~al.}(2023)\citenamefont {Concha}, \citenamefont {Henr{\'\i}quez-Baez}, \citenamefont {Rodr{\'\i}guez},\ and\ \citenamefont {Vera}}]{Concha:2023qcs}%
  \BibitemOpen
  \bibfield  {author} {\bibinfo {author} {\bibfnamefont {P.}~\bibnamefont {Concha}}, \bibinfo {author} {\bibfnamefont {C.}~\bibnamefont {Henr{\'\i}quez-Baez}}, \bibinfo {author} {\bibfnamefont {E.}~\bibnamefont {Rodr{\'\i}guez}}, \ and\ \bibinfo {author} {\bibfnamefont {A.}~\bibnamefont {Vera}},\ }\href {\doibase 10.1103/PhysRevD.107.124043} {\bibfield  {journal} {\bibinfo  {journal} {Phys. Rev. D}\ }\textbf {\bibinfo {volume} {107}},\ \bibinfo {pages} {124043} (\bibinfo {year} {2023})},\ \Eprint {http://arxiv.org/abs/2303.17442} {arXiv:2303.17442 [hep-th]} \BibitemShut {NoStop}%
\bibitem [{\citenamefont {Henr{\'\i}quez-B{\'a}ez}\ \emph {et~al.}(2022)\citenamefont {Henr{\'\i}quez-B{\'a}ez}, \citenamefont {Lagos},\ and\ \citenamefont {Vera}}]{Henriquez-Baez:2022ubu}%
  \BibitemOpen
  \bibfield  {author} {\bibinfo {author} {\bibfnamefont {C.}~\bibnamefont {Henr{\'\i}quez-B{\'a}ez}}, \bibinfo {author} {\bibfnamefont {M.}~\bibnamefont {Lagos}}, \ and\ \bibinfo {author} {\bibfnamefont {A.}~\bibnamefont {Vera}},\ }\href {\doibase 10.1103/PhysRevD.106.064027} {\bibfield  {journal} {\bibinfo  {journal} {Phys. Rev. D}\ }\textbf {\bibinfo {volume} {106}},\ \bibinfo {pages} {064027} (\bibinfo {year} {2022})},\ \Eprint {http://arxiv.org/abs/2208.14239} {arXiv:2208.14239 [hep-th]} \BibitemShut {NoStop}%
\bibitem [{\citenamefont {Flores-Alfonso}(2025)}]{Flores-Alfonso:2024gag}%
  \BibitemOpen
  \bibfield  {author} {\bibinfo {author} {\bibfnamefont {D.}~\bibnamefont {Flores-Alfonso}},\ }\href {\doibase 10.1103/4w6v-4qlv} {\bibfield  {journal} {\bibinfo  {journal} {Phys. Rev. D}\ }\textbf {\bibinfo {volume} {111}},\ \bibinfo {pages} {124014} (\bibinfo {year} {2025})},\ \Eprint {http://arxiv.org/abs/2412.08734} {arXiv:2412.08734 [hep-th]} \BibitemShut {NoStop}%
\bibitem [{\citenamefont {Cacciatori}\ \emph {et~al.}(2021)\citenamefont {Cacciatori}, \citenamefont {Canfora}, \citenamefont {Lagos}, \citenamefont {Muscolino},\ and\ \citenamefont {Vera}}]{Cacciatori:2021neu}%
  \BibitemOpen
  \bibfield  {author} {\bibinfo {author} {\bibfnamefont {S.~L.}\ \bibnamefont {Cacciatori}}, \bibinfo {author} {\bibfnamefont {F.}~\bibnamefont {Canfora}}, \bibinfo {author} {\bibfnamefont {M.}~\bibnamefont {Lagos}}, \bibinfo {author} {\bibfnamefont {F.}~\bibnamefont {Muscolino}}, \ and\ \bibinfo {author} {\bibfnamefont {A.}~\bibnamefont {Vera}},\ }\href {\doibase 10.1007/JHEP12(2021)150} {\bibfield  {journal} {\bibinfo  {journal} {JHEP}\ }\textbf {\bibinfo {volume} {12}},\ \bibinfo {pages} {150} (\bibinfo {year} {2021})},\ \Eprint {http://arxiv.org/abs/2105.10789} {arXiv:2105.10789 [hep-th]} \BibitemShut {NoStop}%
\bibitem [{\citenamefont {Alvarez}\ \emph {et~al.}(2020)\citenamefont {Alvarez}, \citenamefont {Cacciatori}, \citenamefont {Canfora},\ and\ \citenamefont {Cerchiai}}]{Alvarez:2020zui}%
  \BibitemOpen
  \bibfield  {author} {\bibinfo {author} {\bibfnamefont {P.~D.}\ \bibnamefont {Alvarez}}, \bibinfo {author} {\bibfnamefont {S.~L.}\ \bibnamefont {Cacciatori}}, \bibinfo {author} {\bibfnamefont {F.}~\bibnamefont {Canfora}}, \ and\ \bibinfo {author} {\bibfnamefont {B.~L.}\ \bibnamefont {Cerchiai}},\ }\href {\doibase 10.1103/PhysRevD.101.125011} {\bibfield  {journal} {\bibinfo  {journal} {Phys. Rev. D}\ }\textbf {\bibinfo {volume} {101}},\ \bibinfo {pages} {125011} (\bibinfo {year} {2020})},\ \Eprint {http://arxiv.org/abs/2005.11301} {arXiv:2005.11301 [hep-th]} \BibitemShut {NoStop}%
\bibitem [{\citenamefont {Cacciatori}\ \emph {et~al.}(2022)\citenamefont {Cacciatori}, \citenamefont {Canfora}, \citenamefont {Lagos}, \citenamefont {Muscolino},\ and\ \citenamefont {Vera}}]{Cacciatori:2022kag}%
  \BibitemOpen
  \bibfield  {author} {\bibinfo {author} {\bibfnamefont {S.~L.}\ \bibnamefont {Cacciatori}}, \bibinfo {author} {\bibfnamefont {F.}~\bibnamefont {Canfora}}, \bibinfo {author} {\bibfnamefont {M.}~\bibnamefont {Lagos}}, \bibinfo {author} {\bibfnamefont {F.}~\bibnamefont {Muscolino}}, \ and\ \bibinfo {author} {\bibfnamefont {A.}~\bibnamefont {Vera}},\ }\href {\doibase 10.1016/j.nuclphysb.2022.115693} {\bibfield  {journal} {\bibinfo  {journal} {Nucl. Phys. B}\ }\textbf {\bibinfo {volume} {976}},\ \bibinfo {pages} {115693} (\bibinfo {year} {2022})},\ \Eprint {http://arxiv.org/abs/2201.12598} {arXiv:2201.12598 [hep-th]} \BibitemShut {NoStop}%
\bibitem [{\citenamefont {Barlow}\ \emph {et~al.}(2005)\citenamefont {Barlow}, \citenamefont {Doherty},\ and\ \citenamefont {Winstanley}}]{Barlow:2005yd}%
  \BibitemOpen
  \bibfield  {author} {\bibinfo {author} {\bibfnamefont {A.-M.}\ \bibnamefont {Barlow}}, \bibinfo {author} {\bibfnamefont {D.}~\bibnamefont {Doherty}}, \ and\ \bibinfo {author} {\bibfnamefont {E.}~\bibnamefont {Winstanley}},\ }\href {\doibase 10.1103/PhysRevD.72.024008} {\bibfield  {journal} {\bibinfo  {journal} {Phys. Rev. D}\ }\textbf {\bibinfo {volume} {72}},\ \bibinfo {pages} {024008} (\bibinfo {year} {2005})},\ \Eprint {http://arxiv.org/abs/gr-qc/0504087} {arXiv:gr-qc/0504087} \BibitemShut {NoStop}%
\bibitem [{\citenamefont {Harper}\ \emph {et~al.}(2004)\citenamefont {Harper}, \citenamefont {Thomas}, \citenamefont {Winstanley},\ and\ \citenamefont {Young}}]{Harper:2003wt}%
  \BibitemOpen
  \bibfield  {author} {\bibinfo {author} {\bibfnamefont {T.~J.~T.}\ \bibnamefont {Harper}}, \bibinfo {author} {\bibfnamefont {P.~A.}\ \bibnamefont {Thomas}}, \bibinfo {author} {\bibfnamefont {E.}~\bibnamefont {Winstanley}}, \ and\ \bibinfo {author} {\bibfnamefont {P.~M.}\ \bibnamefont {Young}},\ }\href {\doibase 10.1103/PhysRevD.70.064023} {\bibfield  {journal} {\bibinfo  {journal} {Phys. Rev. D}\ }\textbf {\bibinfo {volume} {70}},\ \bibinfo {pages} {064023} (\bibinfo {year} {2004})},\ \Eprint {http://arxiv.org/abs/gr-qc/0312104} {arXiv:gr-qc/0312104} \BibitemShut {NoStop}%
\bibitem [{\citenamefont {Salgado}(2003)}]{Salgado:2003ub}%
  \BibitemOpen
  \bibfield  {author} {\bibinfo {author} {\bibfnamefont {M.}~\bibnamefont {Salgado}},\ }\href {\doibase 10.1088/0264-9381/20/21/003} {\bibfield  {journal} {\bibinfo  {journal} {Class. Quant. Grav.}\ }\textbf {\bibinfo {volume} {20}},\ \bibinfo {pages} {4551} (\bibinfo {year} {2003})},\ \Eprint {http://arxiv.org/abs/gr-qc/0304010} {arXiv:gr-qc/0304010} \BibitemShut {NoStop}%
\bibitem [{\citenamefont {Ashtekar}\ \emph {et~al.}(2003)\citenamefont {Ashtekar}, \citenamefont {Corichi},\ and\ \citenamefont {Sudarsky}}]{Ashtekar:2003jh}%
  \BibitemOpen
  \bibfield  {author} {\bibinfo {author} {\bibfnamefont {A.}~\bibnamefont {Ashtekar}}, \bibinfo {author} {\bibfnamefont {A.}~\bibnamefont {Corichi}}, \ and\ \bibinfo {author} {\bibfnamefont {D.}~\bibnamefont {Sudarsky}},\ }\href {\doibase 10.1088/0264-9381/20/15/310} {\bibfield  {journal} {\bibinfo  {journal} {Class. Quant. Grav.}\ }\textbf {\bibinfo {volume} {20}},\ \bibinfo {pages} {3413} (\bibinfo {year} {2003})},\ \Eprint {http://arxiv.org/abs/gr-qc/0305044} {arXiv:gr-qc/0305044} \BibitemShut {NoStop}%
\bibitem [{\citenamefont {Ashtekar}\ and\ \citenamefont {Corichi}(2003)}]{Ashtekar:2003zx}%
  \BibitemOpen
  \bibfield  {author} {\bibinfo {author} {\bibfnamefont {A.}~\bibnamefont {Ashtekar}}\ and\ \bibinfo {author} {\bibfnamefont {A.}~\bibnamefont {Corichi}},\ }\href {\doibase 10.1088/0264-9381/20/20/310} {\bibfield  {journal} {\bibinfo  {journal} {Class. Quant. Grav.}\ }\textbf {\bibinfo {volume} {20}},\ \bibinfo {pages} {4473} (\bibinfo {year} {2003})},\ \Eprint {http://arxiv.org/abs/gr-qc/0305082} {arXiv:gr-qc/0305082} \BibitemShut {NoStop}%
\bibitem [{\citenamefont {McFadden}\ and\ \citenamefont {Turok}(2005)}]{McFadden:2004ni}%
  \BibitemOpen
  \bibfield  {author} {\bibinfo {author} {\bibfnamefont {P.~L.}\ \bibnamefont {McFadden}}\ and\ \bibinfo {author} {\bibfnamefont {N.~G.}\ \bibnamefont {Turok}},\ }\href {\doibase 10.1103/PhysRevD.71.086004} {\bibfield  {journal} {\bibinfo  {journal} {Phys. Rev. D}\ }\textbf {\bibinfo {volume} {71}},\ \bibinfo {pages} {086004} (\bibinfo {year} {2005})},\ \Eprint {http://arxiv.org/abs/hep-th/0412109} {arXiv:hep-th/0412109} \BibitemShut {NoStop}%
\bibitem [{\citenamefont {Salgado}(2006)}]{Salgado:2005hx}%
  \BibitemOpen
  \bibfield  {author} {\bibinfo {author} {\bibfnamefont {M.}~\bibnamefont {Salgado}},\ }\href {\doibase 10.1088/0264-9381/23/14/010} {\bibfield  {journal} {\bibinfo  {journal} {Class. Quant. Grav.}\ }\textbf {\bibinfo {volume} {23}},\ \bibinfo {pages} {4719} (\bibinfo {year} {2006})},\ \Eprint {http://arxiv.org/abs/gr-qc/0509001} {arXiv:gr-qc/0509001} \BibitemShut {NoStop}%
\bibitem [{\citenamefont {Jackiw}(2006)}]{Jackiw:2005su}%
  \BibitemOpen
  \bibfield  {author} {\bibinfo {author} {\bibfnamefont {R.}~\bibnamefont {Jackiw}},\ }\href {\doibase 10.1007/s11232-006-0090-9} {\bibfield  {journal} {\bibinfo  {journal} {Theor. Math. Phys.}\ }\textbf {\bibinfo {volume} {148}},\ \bibinfo {pages} {941} (\bibinfo {year} {2006})},\ \Eprint {http://arxiv.org/abs/hep-th/0511065} {arXiv:hep-th/0511065} \BibitemShut {NoStop}%
\bibitem [{\citenamefont {Ay{\'o}n-Beato}\ and\ \citenamefont {Hassaine}(2023)}]{Ayon-Beato:2023lrn}%
  \BibitemOpen
  \bibfield  {author} {\bibinfo {author} {\bibfnamefont {E.}~\bibnamefont {Ay{\'o}n-Beato}}\ and\ \bibinfo {author} {\bibfnamefont {M.}~\bibnamefont {Hassaine}},\ }\href {\doibase 10.1016/j.aop.2023.169446} {\bibfield  {journal} {\bibinfo  {journal} {Annals Phys.}\ }\textbf {\bibinfo {volume} {458}},\ \bibinfo {pages} {169446} (\bibinfo {year} {2023})},\ \Eprint {http://arxiv.org/abs/2307.04048} {arXiv:2307.04048 [hep-th]} \BibitemShut {NoStop}%
\bibitem [{\citenamefont {Babichev}\ \emph {et~al.}(2022)\citenamefont {Babichev}, \citenamefont {Charmousis}, \citenamefont {Hassaine},\ and\ \citenamefont {Lecoeur}}]{Babichev:2022awg}%
  \BibitemOpen
  \bibfield  {author} {\bibinfo {author} {\bibfnamefont {E.}~\bibnamefont {Babichev}}, \bibinfo {author} {\bibfnamefont {C.}~\bibnamefont {Charmousis}}, \bibinfo {author} {\bibfnamefont {M.}~\bibnamefont {Hassaine}}, \ and\ \bibinfo {author} {\bibfnamefont {N.}~\bibnamefont {Lecoeur}},\ }\href {\doibase 10.1103/PhysRevD.106.064039} {\bibfield  {journal} {\bibinfo  {journal} {Phys. Rev. D}\ }\textbf {\bibinfo {volume} {106}},\ \bibinfo {pages} {064039} (\bibinfo {year} {2022})},\ \Eprint {http://arxiv.org/abs/2206.11013} {arXiv:2206.11013 [gr-qc]} \BibitemShut {NoStop}%
\bibitem [{\citenamefont {Ay{\'o}n-Beato}\ \emph {et~al.}(2025)\citenamefont {Ay{\'o}n-Beato}, \citenamefont {Hassaine},\ and\ \citenamefont {S{\'a}nchez}}]{Ayon-Beato:2024xgp}%
  \BibitemOpen
  \bibfield  {author} {\bibinfo {author} {\bibfnamefont {E.}~\bibnamefont {Ay{\'o}n-Beato}}, \bibinfo {author} {\bibfnamefont {M.}~\bibnamefont {Hassaine}}, \ and\ \bibinfo {author} {\bibfnamefont {P.~A.}\ \bibnamefont {S{\'a}nchez}},\ }\href {\doibase 10.1140/epjc/s10052-025-14011-5} {\bibfield  {journal} {\bibinfo  {journal} {Eur. Phys. J. C}\ }\textbf {\bibinfo {volume} {85}},\ \bibinfo {pages} {259} (\bibinfo {year} {2025})},\ \Eprint {http://arxiv.org/abs/2408.00086} {arXiv:2408.00086 [hep-th]} \BibitemShut {NoStop}%
\bibitem [{\citenamefont {S{\'a}nchez}(2025)}]{Sanchez:2024xke}%
  \BibitemOpen
  \bibfield  {author} {\bibinfo {author} {\bibfnamefont {P.~A.}\ \bibnamefont {S{\'a}nchez}},\ }\href {\doibase 10.1103/PhysRevD.111.084010} {\bibfield  {journal} {\bibinfo  {journal} {Phys. Rev. D}\ }\textbf {\bibinfo {volume} {111}},\ \bibinfo {pages} {084010} (\bibinfo {year} {2025})},\ \Eprint {http://arxiv.org/abs/2408.07166} {arXiv:2408.07166 [gr-qc]} \BibitemShut {NoStop}%
\bibitem [{\citenamefont {Hernandez-Vera}(2025)}]{Hernandez-Vera:2024zui}%
  \BibitemOpen
  \bibfield  {author} {\bibinfo {author} {\bibfnamefont {U.}~\bibnamefont {Hernandez-Vera}},\ }\href {\doibase 10.1103/PhysRevD.111.084035} {\bibfield  {journal} {\bibinfo  {journal} {Phys. Rev. D}\ }\textbf {\bibinfo {volume} {111}},\ \bibinfo {pages} {084035} (\bibinfo {year} {2025})},\ \Eprint {http://arxiv.org/abs/2412.19388} {arXiv:2412.19388 [hep-th]} \BibitemShut {NoStop}%
\bibitem [{\citenamefont {Fernandes}\ \emph {et~al.}(2022)\citenamefont {Fernandes}, \citenamefont {Carrilho}, \citenamefont {Clifton},\ and\ \citenamefont {Mulryne}}]{Fernandes:2022zrq}%
  \BibitemOpen
  \bibfield  {author} {\bibinfo {author} {\bibfnamefont {P.~G.~S.}\ \bibnamefont {Fernandes}}, \bibinfo {author} {\bibfnamefont {P.}~\bibnamefont {Carrilho}}, \bibinfo {author} {\bibfnamefont {T.}~\bibnamefont {Clifton}}, \ and\ \bibinfo {author} {\bibfnamefont {D.~J.}\ \bibnamefont {Mulryne}},\ }\href {\doibase 10.1088/1361-6382/ac500a} {\bibfield  {journal} {\bibinfo  {journal} {Class. Quant. Grav.}\ }\textbf {\bibinfo {volume} {39}},\ \bibinfo {pages} {063001} (\bibinfo {year} {2022})},\ \Eprint {http://arxiv.org/abs/2202.13908} {arXiv:2202.13908 [gr-qc]} \BibitemShut {NoStop}%
\bibitem [{\citenamefont {Horndeski}(1974)}]{Horndeski:1974wa}%
  \BibitemOpen
  \bibfield  {author} {\bibinfo {author} {\bibfnamefont {G.~W.}\ \bibnamefont {Horndeski}},\ }\href {\doibase 10.1007/BF01807638} {\bibfield  {journal} {\bibinfo  {journal} {Int. J. Theor. Phys.}\ }\textbf {\bibinfo {volume} {10}},\ \bibinfo {pages} {363} (\bibinfo {year} {1974})}\BibitemShut {NoStop}%
\bibitem [{\citenamefont {Babichev}\ \emph {et~al.}(2023{\natexlab{b}})\citenamefont {Babichev}, \citenamefont {Charmousis}, \citenamefont {Hassaine},\ and\ \citenamefont {Lecoeur}}]{Babichev:2023dhs}%
  \BibitemOpen
  \bibfield  {author} {\bibinfo {author} {\bibfnamefont {E.}~\bibnamefont {Babichev}}, \bibinfo {author} {\bibfnamefont {C.}~\bibnamefont {Charmousis}}, \bibinfo {author} {\bibfnamefont {M.}~\bibnamefont {Hassaine}}, \ and\ \bibinfo {author} {\bibfnamefont {N.}~\bibnamefont {Lecoeur}},\ }\href {\doibase 10.1103/PhysRevD.108.024019} {\bibfield  {journal} {\bibinfo  {journal} {Phys. Rev. D}\ }\textbf {\bibinfo {volume} {108}},\ \bibinfo {pages} {024019} (\bibinfo {year} {2023}{\natexlab{b}})},\ \Eprint {http://arxiv.org/abs/2303.04126} {arXiv:2303.04126 [gr-qc]} \BibitemShut {NoStop}%
\bibitem [{\citenamefont {Glavan}\ and\ \citenamefont {Lin}(2020)}]{Glavan:2019inb}%
  \BibitemOpen
  \bibfield  {author} {\bibinfo {author} {\bibfnamefont {D.}~\bibnamefont {Glavan}}\ and\ \bibinfo {author} {\bibfnamefont {C.}~\bibnamefont {Lin}},\ }\href {\doibase 10.1103/PhysRevLett.124.081301} {\bibfield  {journal} {\bibinfo  {journal} {Phys. Rev. Lett.}\ }\textbf {\bibinfo {volume} {124}},\ \bibinfo {pages} {081301} (\bibinfo {year} {2020})},\ \Eprint {http://arxiv.org/abs/1905.03601} {arXiv:1905.03601 [gr-qc]} \BibitemShut {NoStop}%
\bibitem [{\citenamefont {Lu}\ and\ \citenamefont {Pang}(2020)}]{Lu:2020iav}%
  \BibitemOpen
  \bibfield  {author} {\bibinfo {author} {\bibfnamefont {H.}~\bibnamefont {Lu}}\ and\ \bibinfo {author} {\bibfnamefont {Y.}~\bibnamefont {Pang}},\ }\href {\doibase 10.1016/j.physletb.2020.135717} {\bibfield  {journal} {\bibinfo  {journal} {Phys. Lett. B}\ }\textbf {\bibinfo {volume} {809}},\ \bibinfo {pages} {135717} (\bibinfo {year} {2020})},\ \Eprint {http://arxiv.org/abs/2003.11552} {arXiv:2003.11552 [gr-qc]} \BibitemShut {NoStop}%
\bibitem [{\citenamefont {Fernandes}\ \emph {et~al.}(2020)\citenamefont {Fernandes}, \citenamefont {Carrilho}, \citenamefont {Clifton},\ and\ \citenamefont {Mulryne}}]{Fernandes:2020nbq}%
  \BibitemOpen
  \bibfield  {author} {\bibinfo {author} {\bibfnamefont {P.~G.~S.}\ \bibnamefont {Fernandes}}, \bibinfo {author} {\bibfnamefont {P.}~\bibnamefont {Carrilho}}, \bibinfo {author} {\bibfnamefont {T.}~\bibnamefont {Clifton}}, \ and\ \bibinfo {author} {\bibfnamefont {D.~J.}\ \bibnamefont {Mulryne}},\ }\href {\doibase 10.1103/PhysRevD.102.024025} {\bibfield  {journal} {\bibinfo  {journal} {Phys. Rev. D}\ }\textbf {\bibinfo {volume} {102}},\ \bibinfo {pages} {024025} (\bibinfo {year} {2020})},\ \Eprint {http://arxiv.org/abs/2004.08362} {arXiv:2004.08362 [gr-qc]} \BibitemShut {NoStop}%
\bibitem [{\citenamefont {Hennigar}\ \emph {et~al.}(2020)\citenamefont {Hennigar}, \citenamefont {Kubiz{\v{n}}{\'a}k}, \citenamefont {Mann},\ and\ \citenamefont {Pollack}}]{Hennigar:2020lsl}%
  \BibitemOpen
  \bibfield  {author} {\bibinfo {author} {\bibfnamefont {R.~A.}\ \bibnamefont {Hennigar}}, \bibinfo {author} {\bibfnamefont {D.}~\bibnamefont {Kubiz{\v{n}}{\'a}k}}, \bibinfo {author} {\bibfnamefont {R.~B.}\ \bibnamefont {Mann}}, \ and\ \bibinfo {author} {\bibfnamefont {C.}~\bibnamefont {Pollack}},\ }\href {\doibase 10.1007/JHEP07(2020)027} {\bibfield  {journal} {\bibinfo  {journal} {JHEP}\ }\textbf {\bibinfo {volume} {07}},\ \bibinfo {pages} {027} (\bibinfo {year} {2020})},\ \Eprint {http://arxiv.org/abs/2004.09472} {arXiv:2004.09472 [gr-qc]} \BibitemShut {NoStop}%
\bibitem [{\citenamefont {Fernandes}\ \emph {et~al.}(2021)\citenamefont {Fernandes}, \citenamefont {Carrilho}, \citenamefont {Clifton},\ and\ \citenamefont {Mulryne}}]{Fernandes:2021ysi}%
  \BibitemOpen
  \bibfield  {author} {\bibinfo {author} {\bibfnamefont {P.~G.~S.}\ \bibnamefont {Fernandes}}, \bibinfo {author} {\bibfnamefont {P.}~\bibnamefont {Carrilho}}, \bibinfo {author} {\bibfnamefont {T.}~\bibnamefont {Clifton}}, \ and\ \bibinfo {author} {\bibfnamefont {D.~J.}\ \bibnamefont {Mulryne}},\ }\href {\doibase 10.1103/PhysRevD.104.044029} {\bibfield  {journal} {\bibinfo  {journal} {Phys. Rev. D}\ }\textbf {\bibinfo {volume} {104}},\ \bibinfo {pages} {044029} (\bibinfo {year} {2021})},\ \Eprint {http://arxiv.org/abs/2107.00046} {arXiv:2107.00046 [gr-qc]} \BibitemShut {NoStop}%
\bibitem [{\citenamefont {Fernandes}(2020)}]{Fernandes:2020rpa}%
  \BibitemOpen
  \bibfield  {author} {\bibinfo {author} {\bibfnamefont {P.~G.~S.}\ \bibnamefont {Fernandes}},\ }\href {\doibase 10.1016/j.physletb.2020.135468} {\bibfield  {journal} {\bibinfo  {journal} {Phys. Lett. B}\ }\textbf {\bibinfo {volume} {805}},\ \bibinfo {pages} {135468} (\bibinfo {year} {2020})},\ \Eprint {http://arxiv.org/abs/2003.05491} {arXiv:2003.05491 [gr-qc]} \BibitemShut {NoStop}%
\bibitem [{\citenamefont {Hassaine}\ \emph {et~al.}(2023)\citenamefont {Hassaine}, \citenamefont {Hernandez-Vera},\ and\ \citenamefont {Lara-Munoz}}]{Hassaine:2023paj}%
  \BibitemOpen
  \bibfield  {author} {\bibinfo {author} {\bibfnamefont {M.}~\bibnamefont {Hassaine}}, \bibinfo {author} {\bibfnamefont {U.}~\bibnamefont {Hernandez-Vera}}, \ and\ \bibinfo {author} {\bibfnamefont {F.}~\bibnamefont {Lara-Munoz}},\ }\href {\doibase 10.1103/PhysRevD.108.104067} {\bibfield  {journal} {\bibinfo  {journal} {Phys. Rev. D}\ }\textbf {\bibinfo {volume} {108}},\ \bibinfo {pages} {104067} (\bibinfo {year} {2023})},\ \Eprint {http://arxiv.org/abs/2309.03024} {arXiv:2309.03024 [hep-th]} \BibitemShut {NoStop}%
\bibitem [{\citenamefont {Bandos}\ \emph {et~al.}(2020)\citenamefont {Bandos}, \citenamefont {Lechner}, \citenamefont {Sorokin},\ and\ \citenamefont {Townsend}}]{Bandos:2020jsw}%
  \BibitemOpen
  \bibfield  {author} {\bibinfo {author} {\bibfnamefont {I.}~\bibnamefont {Bandos}}, \bibinfo {author} {\bibfnamefont {K.}~\bibnamefont {Lechner}}, \bibinfo {author} {\bibfnamefont {D.}~\bibnamefont {Sorokin}}, \ and\ \bibinfo {author} {\bibfnamefont {P.~K.}\ \bibnamefont {Townsend}},\ }\href {\doibase 10.1103/PhysRevD.102.121703} {\bibfield  {journal} {\bibinfo  {journal} {Phys. Rev. D}\ }\textbf {\bibinfo {volume} {102}},\ \bibinfo {pages} {121703} (\bibinfo {year} {2020})},\ \Eprint {http://arxiv.org/abs/2007.09092} {arXiv:2007.09092 [hep-th]} \BibitemShut {NoStop}%
\bibitem [{\citenamefont {Canfora}\ \emph {et~al.}(2025{\natexlab{b}})\citenamefont {Canfora}, \citenamefont {Corral}, \citenamefont {Diez}, \citenamefont {Guajardo},\ and\ \citenamefont {Oliva}}]{Canfora:2025gwm}%
  \BibitemOpen
  \bibfield  {author} {\bibinfo {author} {\bibfnamefont {F.}~\bibnamefont {Canfora}}, \bibinfo {author} {\bibfnamefont {C.}~\bibnamefont {Corral}}, \bibinfo {author} {\bibfnamefont {B.}~\bibnamefont {Diez}}, \bibinfo {author} {\bibfnamefont {L.}~\bibnamefont {Guajardo}}, \ and\ \bibinfo {author} {\bibfnamefont {J.}~\bibnamefont {Oliva}},\ }\href@noop {} {\  (\bibinfo {year} {2025}{\natexlab{b}})},\ \Eprint {http://arxiv.org/abs/2511.15437} {arXiv:2511.15437 [hep-th]} \BibitemShut {NoStop}%
\bibitem [{\citenamefont {Cirilo-Lombardo}(2023)}]{Cirilo-Lombardo:2023poc}%
  \BibitemOpen
  \bibfield  {author} {\bibinfo {author} {\bibfnamefont {D.~J.}\ \bibnamefont {Cirilo-Lombardo}},\ }\href {\doibase 10.1142/S0219887823502389} {\bibfield  {journal} {\bibinfo  {journal} {Int. J. Geom. Meth. Mod. Phys.}\ }\textbf {\bibinfo {volume} {20}},\ \bibinfo {pages} {2350238} (\bibinfo {year} {2023})}\BibitemShut {NoStop}%
\bibitem [{\citenamefont {Avil{\'e}s}\ \emph {et~al.}(2024)\citenamefont {Avil{\'e}s}, \citenamefont {Corral}, \citenamefont {Izaurieta}, \citenamefont {Valdivia},\ and\ \citenamefont {Vera}}]{Aviles:2024muk}%
  \BibitemOpen
  \bibfield  {author} {\bibinfo {author} {\bibfnamefont {L.}~\bibnamefont {Avil{\'e}s}}, \bibinfo {author} {\bibfnamefont {C.}~\bibnamefont {Corral}}, \bibinfo {author} {\bibfnamefont {F.}~\bibnamefont {Izaurieta}}, \bibinfo {author} {\bibfnamefont {O.}~\bibnamefont {Valdivia}}, \ and\ \bibinfo {author} {\bibfnamefont {C.}~\bibnamefont {Vera}},\ }\href {\doibase 10.1103/PhysRevD.109.084039} {\bibfield  {journal} {\bibinfo  {journal} {Phys. Rev. D}\ }\textbf {\bibinfo {volume} {109}},\ \bibinfo {pages} {084039} (\bibinfo {year} {2024})},\ \Eprint {http://arxiv.org/abs/2402.04503} {arXiv:2402.04503 [gr-qc]} \BibitemShut {NoStop}%
\end{thebibliography}%

\end{document}